# Sub-cycle metrology of bright quantum light

S. Gholam-Mirzaei[1*], M. T. Weil[1,2,4*], D. N. Purschke[1,3], K. M. Kowalczyk[1], A. Staudte[1], D. M. Villeneuve[1], P. B. Corkum[1], J. S. Lundeen[2,4], T. J. Hammond[5], and G. Vampa[1].

[1]Joint Attosecond Science Laboratory (JASLab), National Research Council of Canada and University of Ottawa, Ottawa, ON, Canada

[2]Department of Physics, University of Ottawa, Ottawa ON K1N 6N5, Canada

[3]Laboratory for Laser Energetics, University of Rochester, Rochester NY, 14623-1299, USA

[4]Nexus for Quantum Technologies, University of Ottawa, Ottawa ON K1N 6N5, Canada

[5]Department of Physics, University of Windsor, Windsor ON N9B 3P4, Canada

* These authors contributed equally to this work.

**In quantum optics, quantization of the electromagnetic field typically occurs in a finite volume – a cavity – which results in discrete frequency modes where photons are created, annihilated and exchanged between such modes[1,2]. As a result, evolution of quantum optical states is periodic in the carrier wave of the field, measurement protocols return cycle-averaged information[3], and any sub-cycle evolution that is foundational to many light-matter interactions, especially at high field strengths[4-10], remains hidden. Adapting an attosecond technique[11-13], here we capture the electric-field evolution of a quantum optical state, femtosecond bright squeezed vacuum[14], with sub-cycle precision. We find that it consists of many stochastic, time-localized bursts within each pump pulse whose phase randomly switches between two values. We exploit the random phase flips to generate quantum random bit sequences with a generation rate that can reach petahertz frequencies. In addition, the sub-cycle resolution allows us to measure coherence functions of the waveforms between any two times, which we explain with a superposition of time-limited modes[15]. These results bridge attosecond metrology and quantum optics and pave the way to measuring quantum light-matter interactions as they evolve on a few-femtosecond time scale, and integrate quantum randomness in petahertz electronics[9,10].**

Sub-cycle dynamics characterize many light-matter interactions, especially with intense light, where tunnel or multi-photon ionization initiate high-order harmonic emission[4,5] and light-induced currents in dielectrics[6-8], and enables optoelectronic operations at petahertz rates[9,10]. These strong-field processes have been shown to generate quantum correlations within the matter system[16-18], while some evidence suggests that quantum-optical correlations are created as well[19]. Thus, time-resolved metrology of quantum-optical states with sub-laser cycle temporal resolution is desirable to understanding strong-field quantum-optical interactions.

Sub-cycle quantum-optical metrology was developed in a series of pioneering works, where the vacuum fluctuations at mid-infrared and terahertz frequencies were resolved with sub-cycle precision[20,21] along with their delay-dependent first-order correlation[22]. This is in stark contrast to established measurement methods that integrate over the duration of the temporal modes with a matched local oscillator and hence extract multi-cycle averaged information about the quantum properties of the light[3,23,24]. Theoretical proposals to extend this method, relying on electro-optic sampling, to allow simultaneous measurement of both field quadratures were also put forth[25,26]. These advancements would enable sub-cycle quantum state tomography of any arbitrary state.

Motivated by the need to develop sub-cycle quantum-optical metrology of higher frequency fields typically employed in strong-field interactions, here we extend an attosecond technique, solid-state Tunneling Ionization with a Perturbation for the Time-domain Observation of an Electric field (TIPTOE)[11,12], to measure the electric field of femtosecond bright squeezed vacuum (BSV) as it evolves within each laser pulse[13]. We show that nonlinear photoexcitation in a silicon-based image sensor provides direct time-domain access to the ensemble statistics of stochastic waveforms, which are composed of temporally localized bursts. We characterize the properties of these bursts, extract two-dimensional time-resolved correlation functions, and demonstrate that the phase of each burst assumes one of two values separated by π; i.e., a bimodal distribution characteristic of squeezed vacuum. Assigning a logical value to the quantum random phase of each burst in a laser shot enables extraction of binary sequences that pass standard statistical tests, thus demonstrating a route toward quantum random number generation (QRNG) approaching petahertz bandwidth. The ability to resolve the oscillating electric field of quantum light brings quantum optical phenomena into the sub-cycle time domain, where the quantum behavior of light becomes a directly measurable, high-bandwidth resource for certified randomness generation.

**Results and discussion**

The experiment is sketched in Fig. 1a and detailed in the Methods section. A Bright squeezed vacuum (BSV) beam, centered at a wavelength of 1.6 μm, is generated via high-gain spontaneous parametric down-conversion in two BBO crystals placed in series, pumped by 50 fs pulses at a center wavelength of 800 nm and repetition rate of 1 kHz[14]. The BSV intensity averaged over 1,000 shots is ~0.42 MW/cm², corresponding to an average pulse energy of 1.5 nJ. The histogram of the BSV pulse energies is shown in Fig. 1b, yielding a shot-averaged same-time second-order correlation function[27] $g_{BSV}^{(2)}(0) = 2.08$.

To measure the electric field, the BSV is overlapped spatially and temporally with a gate pulse on a silicon CMOS sensor, with the BSV contributing, on average, ~1% of the total intensity at the detector. The gate is a coherent pulse at the same wavelength, obtained from the degenerate output of a commercial optical parametric amplifier and compressed to 16 fs. Nonlinear absorption of the gate pulse through either multi-photon or tunnelling excitation generates conduction-band carriers (i.e., silicon is transparent at 1.6 μm) in a short temporal window, which is estimated to be 12 fs, or 2.25 cycles (see "nonlinear detection and field mapping" in Methods). The histogram of the nonlinear signal is shown in Fig. 1c, yielding $g_{NL}^{(2)}(0) = 1.01$. The BSV field perturbs the instantaneous absorption rate, producing a phase-sensitive signal that is linear in the BSV electric field. A small crossing angle between the beams introduces a spatially varying delay across the sensor, resulting in vertical interference-like fringes, that directly map the instantaneous electric field of the BSV, $E(t)$, convolved with the duration of the nonlinear signal[11,28], in a single-shot image. The phase of $E(t)$ is measured relative to the phase of the gate. The phase of the gate is locked to the phase of the pump, because the coherent reference is obtained from a self-phase modulation process of the pump. The absolute phase of $E(t)$ can be measured by tagging the carrier-envelope phase of the gate at every shot[11], although this is not done here. In addition, since the setup is not interferometrically stable, two reference beams at 800 nm co-propagating with the BSV and gate pulses, respectively, are interfered in a separate region of the sensor to track the path-length jitter between laser shots. This information is used to assign a consistent phase to each BSV shot, apart from a factor of π radians. To ensure that spatial properties of the wavefronts do not affect the interference pattern, the gate beam is spatially filtered and both beams are collimated. The electric field is retrieved via Fourier-domain filtering of the recorded pattern (see Supplementary Information).

Once $E(t)$ is reconstructed for each shot, we analyze the temporal envelope $|E(t)|^2$. In contrast to coherent laser pulses, BSV does not produce a single deterministic pulse, but rather stochastic emission *bursts* within the pump envelope. These independent bursts arise from amplification of independent vacuum fluctuations. We employ a nonlinear iterative curve-fitting procedure and find that, of the 10,000 shots, 52% are single-burst, 41% exhibit two bursts, and the remainder contain three or more bursts. Representative single- and two-burst waveforms are shown in Fig. 1d, each for 25 shots. The darker solid lines are the mean envelopes of each group. Several features are apparent. First, the nodes of subsequent laser shots are largely fixed in time, while the field amplitude fluctuates randomly (the field amplitude of each shot in the figure is normalized, but the power fluctuations are apparent in Fig. 1b). The fixed-node structure indicates that the observed bursts are phase-locked to the gate, which carries the same phase of the pump. This is because in parametric oscillators the pump selects the quadrature to amplify from an initial vacuum state, that contains all possible phases. As a result, the phase of the squeezed vacuum is linked to half the phase of its pump, up to a sign (a constant factor of π/4 is not measured in the experiment)[29]. The temporal structure of the envelopes is analyzed below.

**Envelope analysis**

In addition to determining the number of bursts within each shot, the nonlinear iterative curve-fitting procedure is used to extract the amplitude, duration, and emission time of the envelopes of such bursts. Figures 2a-c show two-dimensional histograms of the amplitudes of the bursts in the single-burst, two-bursts and three-bursts shots, respectively, as a function of the emission time. A competition between the single- and multi-burst shots is apparent, with single-burst shots being on average stronger and emitted near the center of the emission window (which likely corresponds to the time-dependent gain profile), while multi-burst shots are more likely to be lower in amplitude and emitted near the tails of the emission window. A two-dimensional histogram of the amplitudes of adjacent bursts in the multi-burst shots is shown in Fig. 2d. A small correlation is apparent, with the strongest bursts being associated with the weakest adjacent ones. Supplementary Figure 5b shows a plot of the fitted duration of the bursts. The median temporal width is 17 fs (full-width at half-maximum), longer than the 10 fs estimated from the spectral bandwidth[30], likely due to temporal chirp and slight convolution with the duration of the nonlinear gate. Analysis of emission times (Fig. 2e) reveals a median separation of 26 fs between successive bursts, indicating partial temporal overlap. This overlap region gives rise to interference effects in the reconstructed fields, as discussed below. The minimum separation also explains the amplitude correlation of Fig. 2d, since the strongest bursts are emitted in the middle of the gain window, pushing the adjacent burst to the very tail of the gain window, where the gain is minimal.

Remarkably, the observed temporal structure is consistent with a modal description based on Whittaker–Shannon time-limited modes[15] rather than conventional Schmidt modes. While temporal Schmidt modes are delocalized in time over a duration on the order of the pump pulse, Whittaker-Shannon modes are localized in time with minimal mutual overlap, providing a convenient basis for describing the emission of bursts in femtosecond parametric amplification. The measured burst durations, minimum separations, and coherence functions (see below) quantitatively agree with this time-confined modal structure (see Supplementary Information). Although single-burst[31] and multi-burst[32] emission in BSV has been measured recently, to our knowledge our work constitutes the first application of a temporally localized modal basis to understanding measurements of broadband squeezed vacuum.

The photon-number distribution of all bursts is shown in Fig. 2f. It has a $g_i^{(2)}(0) = 2.94$. The high $g_i^{(2)}(0)$ indicates that the bursts represent independent stochastic degrees of freedom rather than a deterministic substructure of a single extended mode. The stochastic independence of the bursts is further confirmed by the fact that the burst-resolved $g_i^{(2)}(0)$ is significantly larger than the shot-resolved $g_{BSV}^{(2)}(0)$ reported in Fig. 1b. As we show below, the reduction in the shot-resolved $g_{BSV}^{(2)}(0)$ is due to averaging of the independent fluctuations of all the bursts within a laser shot, which decreases the shot-to-shot power fluctuations, akin to how modal averaging affects the $g^{(2)}(0)$ of squeezed vacuum[33]. Finally, given that $g_i^{(2)}(0)$ of the bursts is nearly equal to that of single-mode squeezed vacuum, $g^{(2)}(0) = 3$, these bursts may map to temporally localized modes as suggested by the Whittaker-Shannon decomposition.

The stochastic independence of each temporal burst can be harnessed for extracting quantum information at multi-THz rates, as we show below.

**Quantum random number generation**

Having analyzed the modal structure of BSV, we now turn our attention to the phase of the carrier wave. Squeezed vacuum is known to exhibit a pump-locked phase that randomly flips by $\pi$ radians, producing a bimodal phase distribution that can be binarized into random bits and utilized for Quantum Random Number Generation (QRNG)[29,34]. To demonstrate this concept, we extract the carrier phase of adjacent temporal bursts in the multi-burst shots, $\phi_i$ and $\phi_{i+1}$, using the nonlinear iterative curve-fitting procedure. While the first step of the curve-fitting procedure determines the envelope parameters as discussed above, the second step determines the phases $\phi_i$ and $\phi_{i+1}$ which characterize the carrier oscillation of each burst in the femtosecond BSV field (see Supplementary Information). The resulting histogram of $\Delta\phi = \phi_{i+1} - \phi_i$ (Fig. 3a) is distinctly bimodal, with two equally likely phases separated by π radians; i.e., the field of one burst can randomly flip sign relative to the other. Four randomly selected shots from each of the two clusters are presented in panel b. When the signs of the two bursts are equal ($\Delta\phi = 5.5$ rads or ~7π/4 rads, left panels), their overlap region exhibits constructive interference, producing a continuous carrier oscillation across both bursts despite the envelope displaying two distinct maxima (these shots are indeed not equivalent to single-burst shots). In contrast, when $\Delta\phi = 2.2$ rads (or ~ 3π/4 rads, right panels), the carrier oscillation reverses sign in the overlap region, resulting in a sharp phase discontinuity.

The discrete $\pi$ phase symmetry observed here reflects spontaneous symmetry breaking in parametric amplification and constitutes a quantum source of randomness. This phase randomness has, so far, been demonstrated to occur inside optical parametric oscillators when the pump is periodically reset[29,34], and between laser shots in travelling-wave parametric amplification [31]. The case of shot-to-shot uncorrelated phase symmetry is shown in Fig. 3c where a short pump pulse tends to support amplification of primarily one burst per shot. In contrast, our demonstration supports uncorrelated phase symmetry between temporal bursts separated by only 25 fs within a single pump pulse.

Grouping the phase histogram in two distinct phase classes (red and blue bars in Fig. 3a), logical values can be assigned to the relative random phase of each burst pair in a laser shot, as indicated at the bottom of Fig. 3d. The resulting sequence from the stream of 4,569 bits passes all NIST statistical randomness tests allowed by the length of the sequence[35] (see Supplementary Information). As shown in Fig. 3d, using a longer pump pulse to support the amplification of many localized bursts (i.e., more than two per shot on average) and/or binarizing the absolute phase of each burst rather than the relative phase of adjacent bursts, will allow for the encoding of more than one bit in each laser shot. This capability illustrates how direct time-domain access to quantum phase symmetry can enable randomness extraction at near-PHz rates, thus paving the way for integrating quantum random number generators in PHz electronics[9,10].

**Time-dependent coherence**

Direct access to the electric field $E(t)$ enables evaluation of coherence functions in the time domain, rather than in frequency space or through commonly employed time-integrated measurements. Using the measured waveforms, we define the two-dimensional time-resolved first- and second-order coherence functions[15], $g^{(1)}(t_1,t_2) = \langle E^*(t_1)E(t_2)\rangle/\sqrt{\langle I(t_1)\rangle\langle I(t_2)\rangle}$ and $g^{(2)}(t_1,t_2) = \langle I(t_1)I(t_2)\rangle/[\langle I(t_1)\rangle\langle I(t_2)\rangle]$, directly from the entire dataset (10,000 shots) (see Supplementary Information). Here, $\langle \cdots \rangle$ indicates an average over the shots. $g^{(1)}(t_1,t_2)$ and $g^{(2)}(t_1,t_2)$ are a measure of field and intensity correlations, respectively, between times $t_1$ and $t_2$, which are referenced to the gate field, and thus correspond to two different times in the optical cycle of the BSV. Therefore, these correlation functions provide absolute timing information about the evolution of coherence. As the electric field at any time *t* is known with every laser shot, this measurement technique also provides access to any higher-order correlations $g^{(n)}(t_1,t_2)$ without the need for a separate delay line or additional probe pulses. This would be particularly useful for studying the sub-cycle dynamics of non-Gaussian processes[36,37].

Figure 4a shows $|g^{(1)}(t_1,t_2)|$. The width of the distribution along any vertical or horizontal lineout determines the coherence length, which is ~ 40 fs (FWHM) near the center of the envelope, but increases in its wings. Figure 4c, instead, displays $g^{(2)}(t_1,t_2)$. Near the center of the pulse ($t_1 = t_2 = 0$), the same-time correlation reaches a value of 3, consistent with the intensity fluctuations of single-mode squeezed vacuum. However, it steadily drops to 2 along the same-time diagonal ($t_1 = t_2 \neq 0$); the value commonly associated with thermal noise and two-mode squeezing[1,33]. As we show in the Supplementary Information, this behavior is due to the time-confined burst structure: near the center of the pump pulse, where parametric gain is maximal, a single dominant temporal burst governs the emission, and $g^{(2)}(0,0)$ correspondingly approaches this single-burst value. In the temporal wings of the pump pulse, the gain decreases and additional weaker bursts can be generated. When two statistically independent bursts overlap in time, their intensity fluctuations partially average, reducing $g^{(2)}(t_1 = t_2)$ to 2, as similarly understood for modal averaging[33], and consistent with the time-unresolved measurement reported in Fig. 1b. A similar effect has been measured in the spectral domain, where the contribution of multiple spectral "modes" was shown to decrease $g^{(2)}(\omega)$ in the tails of the spectrum[38]. Because the emission times of the bursts fluctuate randomly from shot to shot, the overlap region shifts across the tails of the pulse, producing a smooth decrease of $g^{(2)}(t_1 = t_2)$ towards 2. The measured lower bound $g^{(2)}(t_1 = t_2)$ $\geq 2$ indicates that, at any given time, at most two bursts significantly overlap. This is consistent with the observed relation between the mean mode duration and their average temporal separation. As presented in Fig. 4b,d, the agreement with the two-time coherence functions expected from a Whittaker-Shannon decomposition is remarkable, despite their analytical form relying on the low-gain parametric down-conversion Hamiltonian (photon pair production regime). Deviations from experiments may be due to experimental imperfections and may also provide an indication of high-gain corrections[39]. The parameters used in the calculations are presented in the Supplementary Information.

**Conclusion**

We have demonstrated single-shot sub-cycle reconstruction of the electric field of a multimode bright squeezed vacuum state using solid-state TIPTOE. The emission is characterized, largely, by statistically independent temporal bursts that are well-localized in time and that behave like single-mode squeezed vacuum. The measurement of temporally localized bursts provides the first application of the recently proposed Whittaker-Shannon decomposition of femtosecond squeezed vacuum. We define and analyze two-dimensional time-dependent coherence functions that reveal a time-resolved structure that is inaccessible to standard time-averaged measurements. Such functions can be used to reveal and control femtosecond to attosecond-fast quantum correlations and statistical properties as they develop in real time during high-field and attosecond interactions in matter.

The stochastic localized waveforms reveal a bimodal phase-difference distribution characteristic of squeezed vacuum that we exploited to generate a stream of quantum random bits. Combining femtosecond squeezed light sources with high temporal-mode capacity (i.e.

longer pump pulse durations) and attosecond-precision field sampling extends this concept toward scalable QRNG approaching petahertz rates. This direction connects attosecond metrology to quantum information science and establishes a platform where the quantum behavior of light becomes a directly measurable, high-bandwidth resource for certified randomness generation. The convenient use of standard image sensors allows PHz-QRNG at telecom wavelengths.

## Materials and Methods

Experimental setup

A schematic of the experimental setup is shown in Extended Data Fig. 1. A Ti:sapphire regenerative amplifier (Coherent Legend) delivers pulses centered at 800 nm with a duration of 50 fs (FWHM) at a repetition rate of 1 kHz. Approximately 1.1 mJ of pulse energy is used to pump a commercial optical parametric amplifier (Light Conversion TOPAS).

In the degenerate OPA configuration, cross-polarized signal and idler beams are generated at 1.6 µm. The signal beam is selected as the coherent arm in TIPTOE using a wire-grid polarizer and subsequently spectrally broadened by focusing into a 2 mm-thick YAG plate. The central portion of the broadened spectrum is selected using a spatial filtering aperture.

Approximately 30 µJ of pump pulse energy is directed to the generation of bright squeezed vacuum (BSV) centered at 1.6 µm via high-gain spontaneous parametric down-conversion in a Type-I phase-matching geometry. The nonlinear medium consists of two 2 mm-thick, uncoated β-barium borate (BBO) crystals arranged in tandem, separated by ~20 cm. Prior to the crystals, the pump beam is telescoped to a beam diameter of approximately 0.5 mm to achieve the desired intensity for high parametric gain. A variable neutral density filter is used to control the pump energy incident on the crystals. Approximately 1.5 nJ of average BSV energy was used for TIPTOE. The tandem configuration of the BBO crystals serves not only to increase gain, but also to restrict the spatial and spectral mode content of the emitted radiation[14]. In particular, the second crystal preferentially amplifies a subset of the spatial modes generated in the first crystal, resulting in partial mode filtering.

The BSV and coherent reference beams are combined non-collinearly with a small crossing angle ($\theta < 6°$) in the horizontal plane. This geometry produces a spatially varying time delay $\tau(x)$ across the detector, given by $\tau(x) \approx (x/c)\sin(\theta)$, where $x$ is the horizontal coordinate on the sensor and $c$ is the speed of light. This mapping enables single-shot acquisition of the temporal waveform E(t) along the spatial axis of the camera. The beams are weakly focused onto a silicon CMOS sensor (FLIR Blackfly BFS-U3-28S5M-C) to ensure a nearly uniform intensity distribution across the interaction region while maintaining sufficient peak intensity for nonlinear excitation. At the CMOS sensor, the BSV beam is ~ 3.3 mm in diameter ($1/e^2$). This is extracted from the vertical width of the envelope of the TIPTOE fringes shown in Supplementary Figure S2b (see "isolating the BSV signal" in the Supplementary Information).

The sub-cycle path-length jitter between the coherent reference and the BSV arms is measured at each shot by sending 800 nm reference signals from each arm onto a separate area of the detector. In the coherent reference arm, this reference beam is obtained by frequency doubling the broadened OPA signal beam in a 50 µm-thick BBO crystal. In the BSV arm, this is the residual BSV pump. The 1.6 µm and 800 nm signals are split by a reflective neutral density filter (OD 0.02) after they are combined on a common path, ensuring that the crossing angles of the two 1.6 µm beams (black lines, TIPTOE path) and of the two 800 nm beams (red lines, references path) are identical. A bandpass filter at 800 nm in the reference arm and a long-pass filter at 1.6 µm in the TIPTOE arm ensure proper separation of the two wavelengths in these arms. Because the BSV pump is much brighter than the frequency doubled coherent reference, a variable neutral density filter wheel (OD 2) is placed on the common reference path, sending the former beam through the high density region of the wheel, and allowing the latter through the transparent region. The interference of the two reference beams yields vertical fringes whose phase is used during the data analysis to correct for the path-length jitter. Because the jitter is measured with 800 nm wavelength, and that a path-length shift of an integer multiple of a wavelength is indistinguishable from no difference at all, the jitter is known only up to a half-wavelength of the BSV. Thus, all BSV fields *of different laser shots* are known up to a sign. However, within the same laser shot, the relative phase between the temporal bursts is a legitimate observable as it is unaffected by path-length jitter. Because of the identical crossing angle to the TIPTOE beams, the fringe spacing is half that of the TIPTOE signal.

Nonlinear detection and field mapping

Silicon is transparent at λ = 1.6 μm. Carrier excitation therefore occurs via a nonlinear process, with the excitation rate depending nonlinearly and monotonically on the instantaneous electric field of the incident light. In the absence of the BSV, the coherent reference alone produces a nonlinear signal that varies slowly in space, according to its slowly varying envelope. When the BSV is present, its electric field interferes with the reference field, modifying the instantaneous field amplitude $E_{total}(t) = E_{ref}(t) + E_{BSV}(t)$.

Because the excitation rate depends nonlinearly on $|E_{total}(t)|$, the presence of the BSV modulates the carrier generation rate in a manner sensitive to both the amplitude and phase of $E_{BSV}(t)$. When the nonlinear excitation is confined to a time window shorter than one optical cycle of the carrier wave, this modulation provides a direct mapping of the instantaneous electric field of the BSV onto the measured signal. These general principles of TIPTOE have been described elsewhere[11,12].

The noncollinear geometry combined with the nonlinear detection results in vertical interference-like fringes on the camera. These fringes correspond to oscillations of the instantaneous electric field of the BSV as a function of delay. Each camera frame therefore contains the full temporal waveform $E_{BSV}(t)$ of a single realization of the squeezed vacuum field.

Stability and temporal resolution

The coherent reference and BSV share the same absolute phase. The coherent reference is the signal beam of a white-light seeded OPA. Thus, provided the white-light generation is stable, $\phi_s = \phi_{800}$, apart from a constant factor. The BSV has the phase of the pump, apart from a random sign and a constant factor of π/4[1,29]. Thus, passive phase stability between the fields is ensured. Interferometric stability of the setup, on the other hand, is independently verified as described above.

The temporal resolution of the measurement, which determines the minimum burst duration that can be measured, is given by the effective duration of the nonlinear excitation window, governed by both the reference pulse duration and the nonlinear response. Nonlinear absorption of the 16 fs reference pulse is measured to scale with $I^{1.7}$ ($I$ is the intensity of the reference on the camera, and 1.7 is the measured order of nonlinearity). Therefore, the effective gating window is estimated to be ~ 12 fs.

**Data availability:** All data used in the analysis is available from the authors upon reasonable request.

**Code availability:** All code used in the analysis is available from the authors upon reasonable request.

**Acknowledgments:** The authors acknowledge G. Thekkadath, F. Bouchard, D. England, P. Bustard and B. Sussman for insightful discussions, and D. Crane and R. Kroeker for technical support.

**Funding:** S.G.-M and P.B.C acknowledge support from Army Research Office through grant No. W911NF-24-1-0214. G.V. acknowledges support from the Quantum Sensing Program of the National Research Council Canada, and from the Natural Sciences and Engineering Research Council through grant RGPIN-2021-04286. G.V., P.B.C., A.S. and J.L. acknowledge support from the Joint Center for Extreme Photonics (JCEP). M.W. acknowledges support from a Canada Graduate Research Scholarship – Doctoral (NSERC). T. J. H acknowledges support from the Natural Sciences and Engineering Research Council grant RGPIN-2019-06877. D.N.P. acknowledges support from the Natural Sciences and Engineering Research Council postdoctoral fellowship (PDF - 578397 – 2023).

**Author contributions:**

Conceptualization: SGM, MW, DNP, GV

Methodology: SGM, MW, TJH, KK, DNP, GV

Investigation: SGM, MW, TJH, GV

Formal Analysis: MW, GV, KK, DNP, SGM

Visualization: MW

Funding acquisition: AS, PBC, JL, GV

Supervision: AS, PBC, JL, GV

Writing – original draft: SGM, MW, GV

Writing – review & editing: SGM, MW, TJH, KK, DNP, AS, DV, PBC, JL, GV

**Competing interests:** Authors declare that they have no competing interests.

**Materials & Correspondence:** Correspondence and material requests should be addressed to G. V. (gvampa@uottawa.ca).

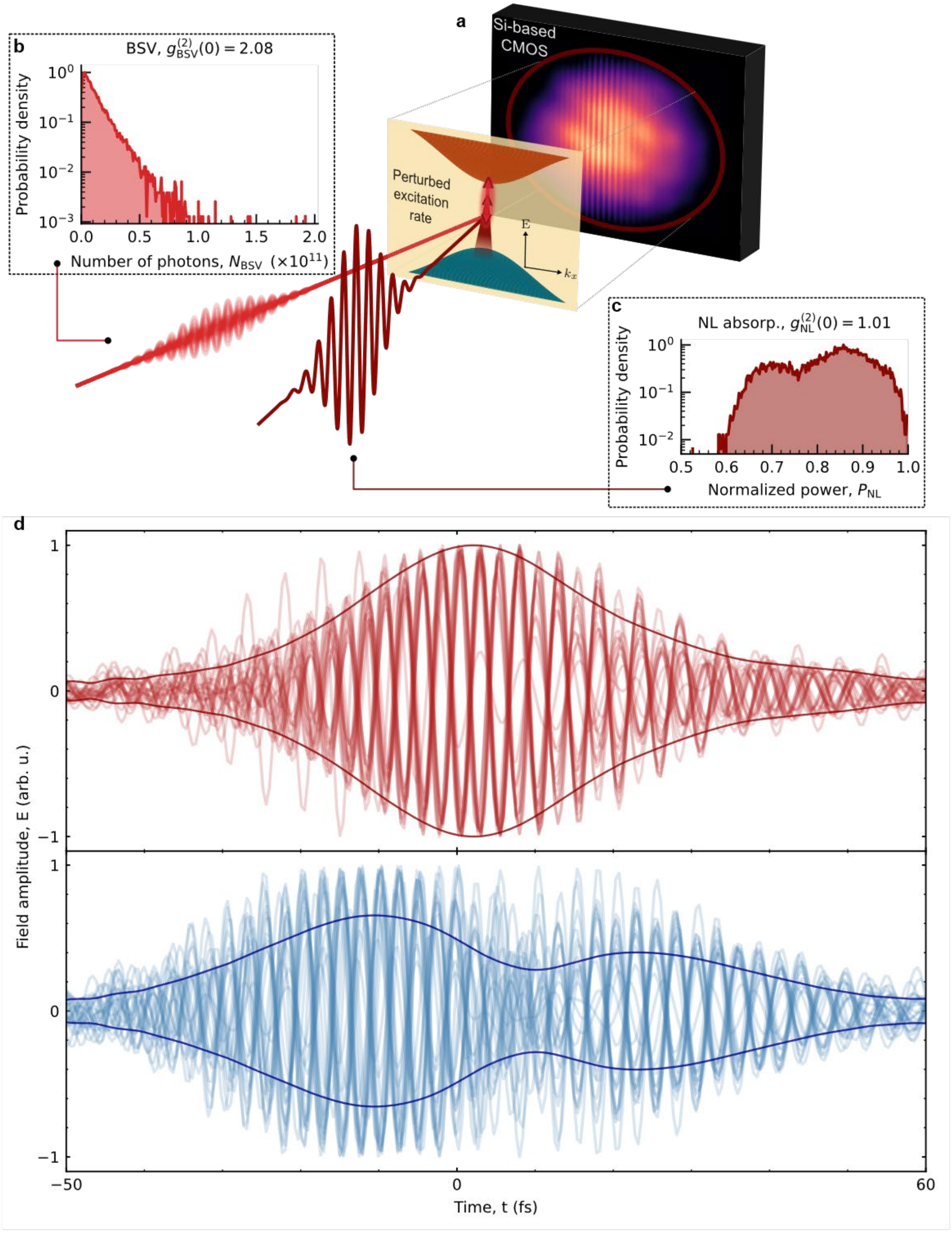
a
Si-based
CMOS
Perturbed
excitation
rate
b
BSV, $g^{(2)}_{\mathrm{BSV}}(0) = 2.08$
Probability density
Number of photons, $N_{\mathrm{BSV}}$ ($\times 10^{11}$)
c
NL absorp., $g^{(2)}_{\mathrm{NL}}(0) = 1.01$
Probability density
Normalized power, $P_{\mathrm{NL}}$
d
Field amplitude, E (arb. u.)
Time, t (fs)

**Fig. 1: TIPTOE measurement of femtosecond Bright-Squeezed Vacuum.**
**a,** Schematics of the experiment. A BSV (blurred, light red line) perturbs the instantaneous absorption rate of a coherent gate pulse (solid, dark red line), increasing or decreasing carrier generation depending on the relative phase of the two fields. The two beams interfere at a small crossing angle on a Si-based CMOS detector generating a horizontal interference pattern that nearly directly maps the time-dependent BSV electric field of each laser shot. **b,** Photon number histogram of the femtosecond BSV showing the characteristic heavy-tail distribution. Photon numbers are estimated by equating the mean measured BSV power and the mean time-integrated intensity of each TIPTOE shot. **c,** The normalized power histogram of the nonlinear absorption signal from the gate pulse alone. The zero-time (average) second-order coherence function value ($g^{(2)}(0)$) for both beams is listed above their respective distributions. **d,** The top panel shows the electric field of 25 shots of single-burst BSV overlayed in-time, as measured with TIPTOE. The bottom panel shows 25 shots of multi-burst BSV overlayed in-time. The solid curves represent the mean envelope of the two sets of measured fields. The envelopes are shifted to the mean emission time of single-burst shots. The phase is relative to the phase of the gate.

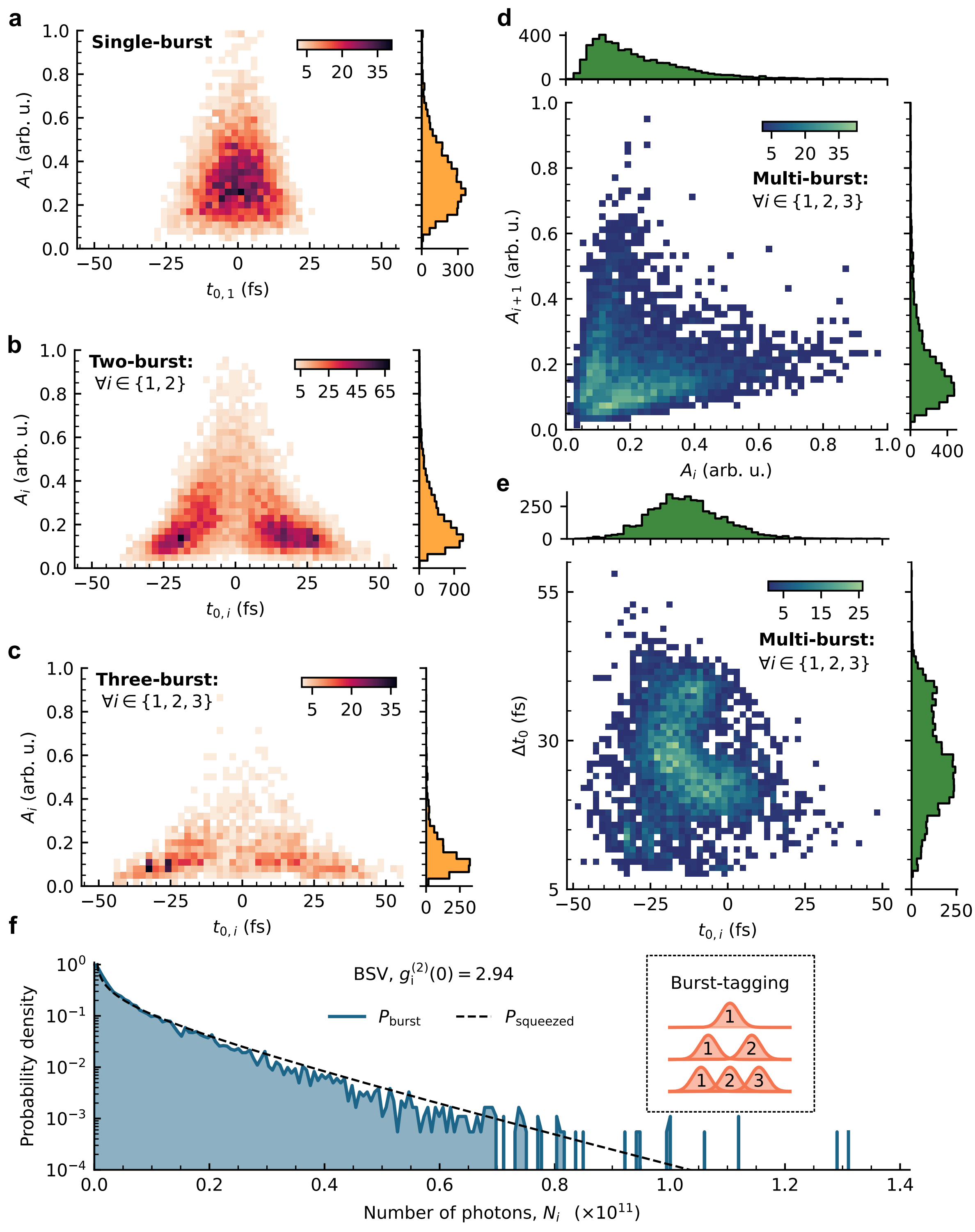

**Fig. 2: Modal analysis of BSV.**
**a-c,** Amplitude vs. emission time distributions for single-burst, two-burst, and three-burst BSV, respectively. **d,** Distribution of the amplitudes of adjacent burst pairs. **e,** Distribution of emission times of adjacent burst pairs (i.e., bursts *i* and *i+1* of the *i*-th burst pair in each shot). The emission time of the first burst in each pair ($t_{0,i}$) is plotted against the time-delay of each pair ($\Delta t_0 = t_{0,i+1} - t_{0,i}$). The marginal distribution along each axis of the histograms is shown in orange in **a-c** and in green in **d, e** on the top and right of each panel. **f,** The burst-resolved photon number distribution of both single and multi-burst BSV as measured via TIPTOE. The dashed curve depicts the expected photon number distribution for single-mode BSV[40]. The zero-time burst-average second-order coherence function value is shown at the top of the distribution. The inset illustrates how each burst is labelled within a given shot. The distributions shown consider the shots whose fits were deemed successful by the iterative curve-fitting procedure (see Supplementary Information).

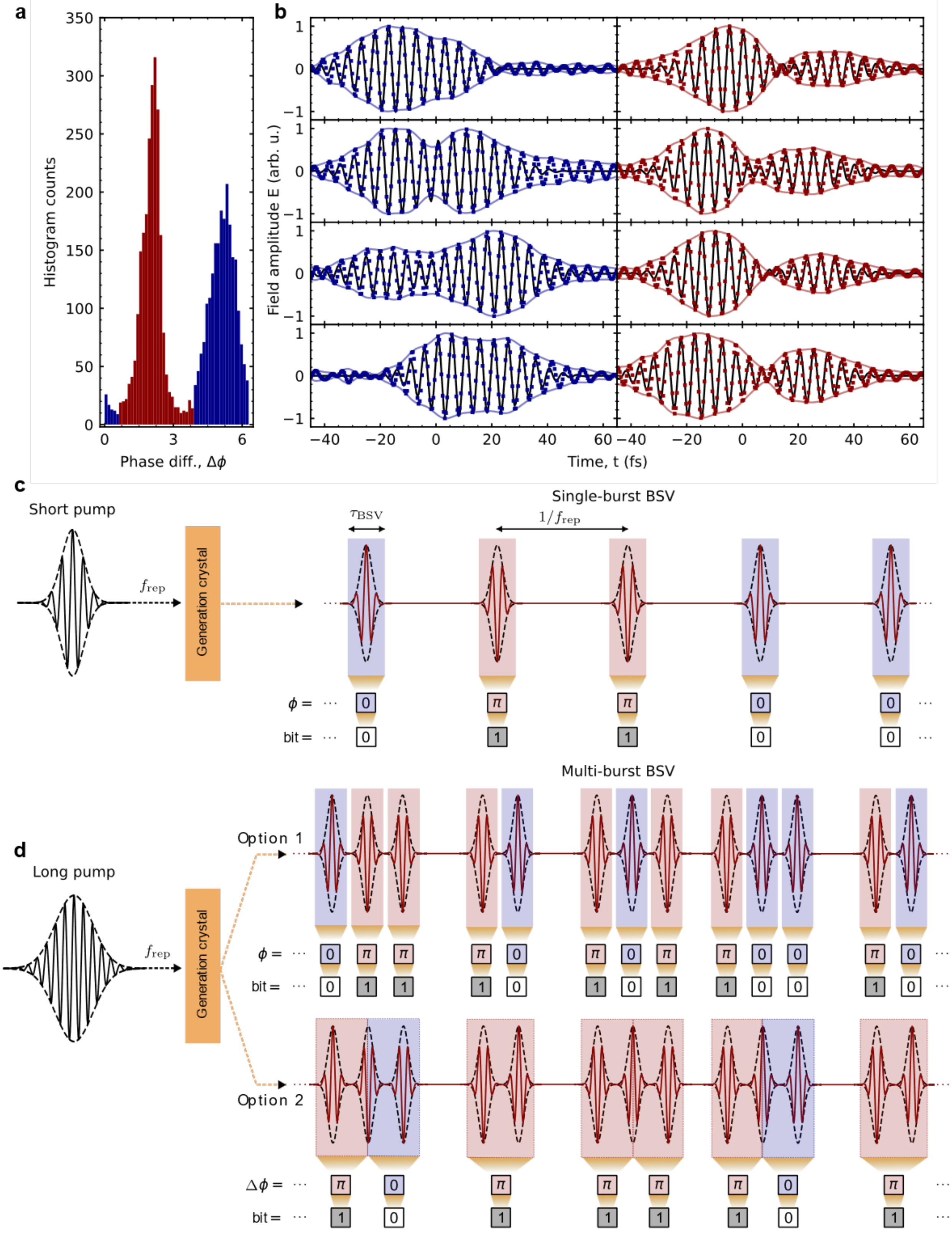
a
b
c
d
Histogram counts
Phase diff., $\Delta\phi$
Field amplitude E (arb. u.)
Time, t (fs)
Short pump
Long pump
$f_{\text{rep}}$
Generation crystal
Single-burst BSV
Multi-burst BSV
$\tau_{\text{BSV}}$
$1/f_{\text{rep}}$
$\phi =$
bit =
$\Delta\phi =$
Option 1
Option 2

**Fig. 3: Binarization of BSV bursts phase difference.**

**a,** The bi-modal histogram of the phase difference between adjacent bursts, $\Delta\phi$. The mean error on the fitted phase-difference is 0.04 rads for the burst-pairs in the red peak and is 0.03 rads for those in the blue peak, and thus the finite width of either peak in the bi-modal distribution may be physical in origin. The red and blue peaks are located at approximately 2.2 rads and 2.2 + π rads, respectively. **b,** Four representative shots of BSV from the blue (red) peak in the bi-modal distribution of panel A are shown in the left (right) panel. The markers correspond to experimentally measured data, the solid blue and red curves represent the envelope of the measured field, and the black curve depicts the results of the curve-fitting procedure. Recall that the sign of the amplitude is not retrieved in the measurement. Only one possible sign for the shot is shown for clarity. **c,** BSV generated from a temporally short pump will primarily exhibit a single burst. The carrier-phase is locked to that of the pump, up to a sign-flip. **d,** A longer pump pulse supports multi-burst BSV. Each of the bursts maintains a phase relationship with the pump, as in the single-burst case. Binarization of multi-burst BSV can be done using either the phase of each burst (top) or using the inter-burst phase difference (bottom). Measuring the phase from a stream of many-burst BSV will allow for random number generation at unprecedented ~ 200 THz rates, given by the duration of the bursts. Note that the inter-burst phase difference in **d** is shown to be exactly π rads or 2π (0) rads for clarity.

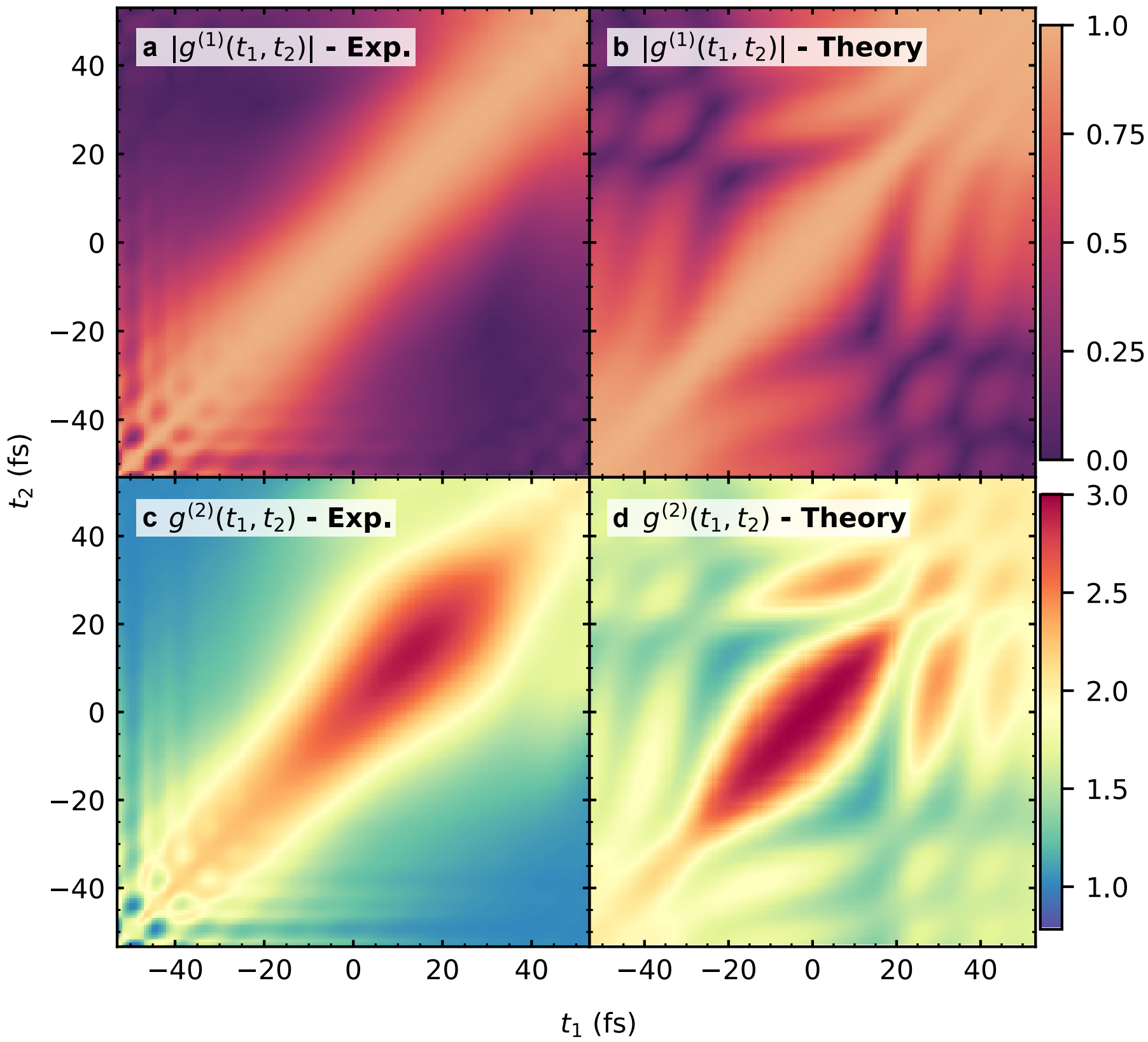

**Fig. 4: Time-resolved quantum coherence functions.**
**a,** The absolute value of the measured first-order coherence function $|g^{(1)}(t_1, t_2)|$ of the generated femtosecond BSV. **b,** The theoretical $|g^{(1)}(t_1, t_2)|$ calculated using the Whittaker-Shannon decomposition. **c,** The measured second-order coherence function $g^{(2)}(t_1, t_2)$ of the generated BSV. **d,** The theoretical $g^{(2)}(t_1, t_2)$ calculated using the Whittaker-Shannon decomposition.

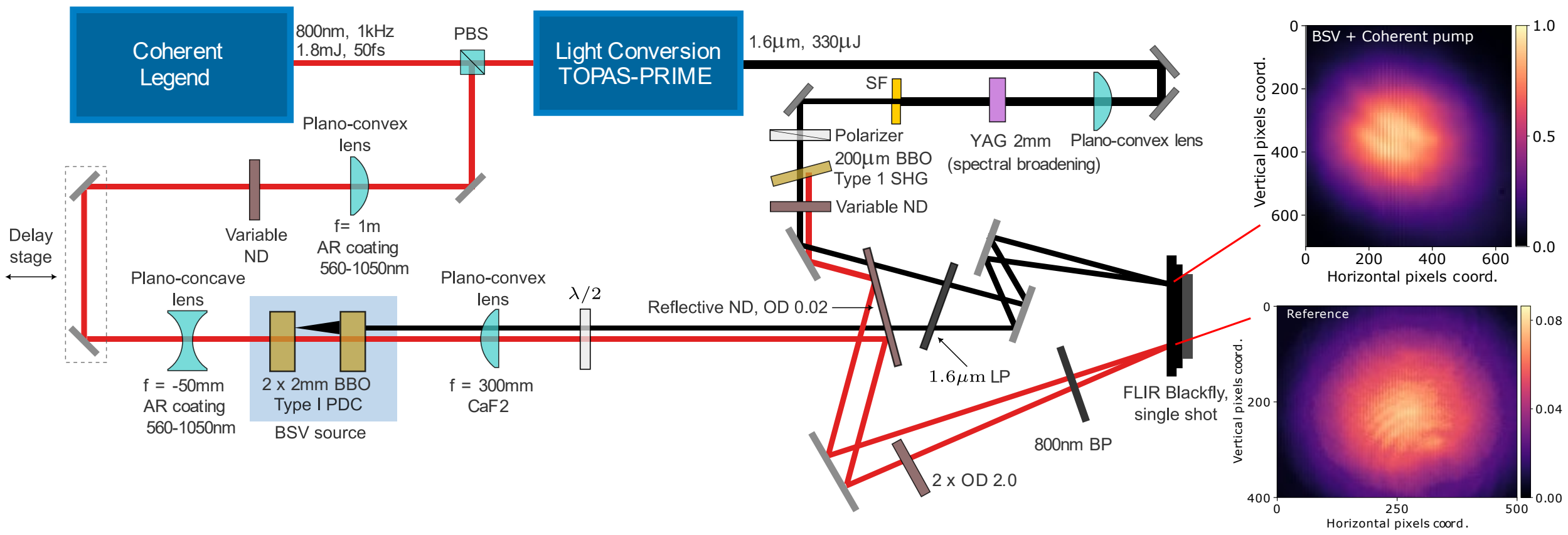

**Extended Data Fig. 1: Extended experimental setup.**
PBS: polarizing beam splitter, SHG: second harmonic generation, ND: neutral density filter, f: focal length, PDC: parametric down conversion, BSV: bright squeezed vacuum, OD: optical density filter, LP: long pass filter, BP: band pass filter.

# Supplementary Information for

## Sub-cycle metrology of bright quantum light

S. Gholam-Mirzaei[1*], M. T. Weil[1,2,4*], D. N. Purschke[1,3], K. M. Kowalczyk[1], A. Staudte[1], D. M. Villeneuve[1], P. B. Corkum[1], J. S. Lundeen[2,4], T. J. Hammond[5], and G. Vampa[1].

[1]Joint Attosecond Science Laboratory (JASLab), National Research Council of Canada and University of Ottawa, Ottawa, ON, Canada

[2]Department of Physics, University of Ottawa, Ottawa ON K1N 6N5, Canada

[3]Laboratory for Laser Energetics, University of Rochester, Rochester NY, 14623-1299, USA

[4]Nexus for Quantum Technologies, University of Ottawa, Ottawa ON K1N 6N5, Canada

[5]Department of Physics, University of Windsor, Windsor ON N9B 3P4, Canada

* These authors contributed equally to this work.

Corresponding author: gvampa@uottawa.ca

This file contains:

Supplementary Discussion:

Isolating the BSV signal

The real-space interference of the 1.6 $\mu$m coherent pump and the weak BSV perturbation, as well as the interference of the 800 nm reference beams is captured directly on the Si-based CMOS sensor. In its raw form, the signal strength is in analog-to-digital units and the axes defining the dimensions of the interference are the pixel coordinates. In Fig. S1a, the raw signal collected on the detector according to the methods described in the previous section is shown. The 1.6 $\mu$m interference pattern is shown in the upper-left and, as expected, the real-space fringe spacing is twice that of the 800 nm interference pattern – as depicted by the two inset panels of the figure.

The first step to extracting $E_{BSV}(t)$ is to perform a Fourier transform of the raw signal along the x-direction (this represents the time axis). The result of this 1D transformation is shown in Fig. S1b where we find five distinct vertical bands corresponding to the DC offset (white region), the 1.6 $\mu$m fringes (positive frequency signal highlighted by the blue region), and the 800 nm fringes (positive frequency signal highlighted by the green region). We note that the elongated horizontal features seen in the 1D transformation correspond to damage spots and/or diffraction from dust particles.

We then perform an inverse Fourier transformation (IFT, again, along the horizontal axis) of each of the three regions of interest (ROI) shown in Fig. S1b, the results of which are shown in Fig. S2, a-c. Panel a corresponds to the IFT of the white ROI and represents the real-space *envelope* of the nonlinear absorption from the coherent reference (left beam) and of linear absorption of both reference beams (right beam). Performing an integral over, for example, the Gaussian-like feature in the upper-left of panel a allows one to track the fluctuations of the nonlinear absorption signal, which is related to the OPA coherent pump. In contrast, panels b and c correspond to the real part of the IFT of the blue and green ROIs, respectively. The real part of $E_{BSV}(t)$ can be *read* directly from the fringes shown in panel b. In combination with the imaginary part of the IFT (not shown in Fig. S2), the full complex BSV waveform is obtained. The BSV signal used in the following analysis is the average over the vertical axis of a region of high-visibility fringes found from such IFT (see the orange region depicted in panel b). An analogous procedure is used for extracting the complex reference signal. In this case, a lineout is used (see the yellow line in panel c). The reconstructed field is obtained up to an overall scaling factor set by the reference field amplitude. Absolute field calibration can in principle be achieved through independent characterization of the nonlinear response function of the detector but is not required for extracting phase statistics and correlation functions.

We calibrate the time axis according to the spacing between the positive and negative frequency peaks in the spectrum of the 800 nm reference signal, as presented in Fig. S2d. Since the fringe spacing corresponds to one period of 800 nm, the two peaks (orange 'x' markers in panel d) are separated in angular frequency by $2 \times \omega_{800}$. The step size in the time-domain can be defined using the product of the number of points between such peaks and this angular frequency spacing [rad/s], divided by 2π rad. Using the calibrated time axis, we can then study the temporal structure of the BSV field. A sample shot of this measured field is shown in panel e.

For further analysis, each BSV field is windowed to remove the ringing at either end of the time axis. This is expected when transforming between truncated spaces, as we do in processing the raw TIPTOE signals. The BSV field is also normalized to the envelope of the OPA pump-induced nonlinearity which is extracted from a lineout of the corresponding region in the IFT of the DC offset (see, for example, the upper-left region of panel a). The time-axis for all figures in the main text and in the Supplementary Information is shifted by the mean emission time found from the single-burst BSV analysis as described in the following subsections.

Correcting the BSV phase

Correction for the relative phase of the BSV between shots using the measured reference signal is used at this stage. For every shot, we subtract half the phase of the pump measured at the positive spatial-frequency peak from the complex BSV field (the factor of two comes from the ratio of the known frequencies of the pump and PDC). This correction ensures that we may perform shot-to-shot comparisons of the measured BSV field without seeing the effects of the overall phase noise to which the entire setup is subject, up to a sign.

Curve-fitting procedure

To study the temporal structure of the measured BSV fields, we employ a nonlinear iterative curve-fitting procedure analogous to that used in standard signal processing for extracting an unknown number of signal peaks, of a given shape, from the overall spectral data[1]. Here, the signal peaks for the measured BSV field manifest as localized bursts within each shot, as introduced in the manuscript. As we are interested in studying not only the local (i.e., burst-resolved) structure of the envelope of the BSV field, but also the carrier oscillation, we require a slightly modified approach. The general idea is as follows: given a measured BSV field, we can first decompose its envelope into the sum of some number of Gaussians and then fit the carrier phase to the real part of the waveform assuming the number and position of the underlying Gaussian bursts are fixed. While this seems straightforward, the number of underlying Gaussian bursts, their overlap, and the relative phase between them complicates the process. It is for this reason that we rely on a rather involved, and iterative scheme as described in the steps below.

*Step 1: Fitting the envelope.* For a given BSV field, the intensity envelope is fit using a model sum of $n$ Gaussian envelopes. For example, for $n$ bursts, the fit model is explicitly the following,

$$y(t) = \sum_{i}^{n} A_i^2 \exp\left[-2\ln(2)\frac{\left(t - t_{0,i}\right)^2}{\sigma_i^2}\right], \tag{1}$$

where the fit parameters $A_i$, $t_{0,i}$, and $\sigma_i$ are, respectively, the amplitude, emission time, and half-width at half-maximum of the electric field of the $i$-th burst. The FWHM of the envelope reported in the main text and in Fig. S4, S5, and S6 corresponds to $\sigma_{I,i} = \sqrt{2}\sigma_i$.

*Step 2: Fitting the phase.* Using the results of step 1, the real part of the BSV field can be fit to a superposition of sinusoids to extract the burst-dependent carrier phase relationships. For example, for $n$ bursts, the fit model is explicitly the following,

$$y(t) = \sum_{i}^{n} A_i \exp\left[-\ln(2)\frac{\left(t - t_{0,i}\right)^2}{\sigma_i^2}\right]\cos(\omega_0 t + Ct^2 + \phi_i)\,. \tag{2}$$

During this step, the number of underlying bursts and their precise emission times $t_{0,i}$ are fixed from the output of step 1. The amplitudes, $\{A_i\}$, and widths, $\{\sigma_i\}$, are given relatively tight bounds of $\pm 15\%$ with respect to the values found in step 1. The *breathing room* given to the amplitude and width parameters ensures that the regions of overlap between multiple bursts can be properly fit. Parameter $C$ allows for a small chirp and is fixed, after an initial calibration procedure (see below). The center frequency $\omega_0$ is also fixed, after an initial calibration procedure. The parameters of interest in this step are the bursts' absolute phases, $\{\phi_i\}$.

*Step 3*: *Evaluation of the fit error.* The mean fit error (MFE) is the root-mean-squared-error evaluated using the difference between the resulting fit of step 2 and the real part of the given BSV field, expressed as a percentage of the amplitude of the BSV envelope. Using this metric, we can compare the performance of the first two steps for a given fit model containing $n$ underlying bursts[1].

The iterative nature of the fit procedure is realized by looping over the steps 1 and 2 until the change in the MFE is less than 0.5%. After each iteration of steps 1 and 2, the number of underlying bursts in the fit model increases by one. The threshold on the change in MFE is used to ensure that our fitting procedure does not overfit the measured field. In other words, the addition of another burst in the fit model is only justified if it decreases the MFE by more than 0.5%. After each iteration, we also verify that no individual burst is located within one HWHM (i.e., $\sigma_i$) of another burst in the final fit. This is to ensure that the algorithm does not attempt to overlay two bursts exactly on top of one another.

The final fit for each BSV field is then the fit produced by the last iteration which satisfied these two conditions. At the end of the procedure, 5163 shots are found to have one burst, 4097 shots are found to have two bursts, and 740 shots are found to have three-or-more bursts. After thresholding the single-burst and multi-burst sets based on MFE (see more below), we find 3916 successful single-burst fits and 3881 successful multi-burst fits. The 3881 successful multi-burst fits correspond to 8450 individual bursts.

*Calibrating the center frequency and the chirp.* We calibrate the center frequency $\omega_0$ of our fitting algorithm by performing an initial two-step least-squares fit on a subset of single-burst shots. For this initial calibration step, we run all the BSV shots through a simple peak-finding algorithm to filter out those with only one peak in the envelope. The height threshold for the peak-finding algorithm is set to a low value such that even a small, secondary peak will result in a shot being labelled as multi-burst. This ensures that the remaining subset of single-burst shots, as found by this initial step, can be fit directly to a sinusoidal waveform with a single-peak Gaussian envelope. This subset of approximately single-burst shots is then fed into our iterative algorithm (although, in this case, there is only one iteration) with the center frequency as a fit parameter.

The result of this calibration step is a distribution of center frequencies from the fits of this subset of single-burst BSV shots. The median value of this distribution is then used as the fixed center frequency for the full curve-fitting procedure. The fixed parameter value is $\omega_0 \sim 1.27 \text{ rad/fs}$. It is crucial for the center frequency to be fixed to avoid coupling with the absolute phases during step 2. A mean error of $0.02 \text{ rad/fs}$ is found for the center frequency fit parameter.

While a simple peak-finding algorithm can be used to identify the number of underlying bursts in some cases (e.g., when the bursts are well-separated), we have found that it is insufficient in cases when there is significant burst overlap and/or when the underlying bursts do not exhibit a stark phase shift between one another. For this reason, *all* the BSV shots collected (i.e., even those which were initially identified as single-burst by the peak-finding algorithm) are analyzed using the iterative curve-fitting procedure following this calibration step.

Finally, an analogous calibration step is used to identify the fixed chirp value to be used in the iterative curve-fitting procedure. The fixed parameter value is $C \sim -1 \times 10^{-5} \text{ fs}^{-2}$ which indicates that there is a low level of chirp in our measured BSV field. This low amount of chirp is expected from the optical setup. The mean error of $5 \times 10^{-5} \text{ fs}^{-2}$ is found for the chirp fit parameter, thus indicating that the fit performance is rather insensitive to the precise value of $C$, so long that the value is near zero.

*General settings.* The method used for performing the nonlinear least-squares fit at each step of the above procedure is Trust Region Reflective[2]. The maximum number of function evaluations used during each nonlinear least-squares fit is $10000 \times n$ where $n$ is the number of underlying bursts in the fit model. Global search techniques such as differential evolution methods[3] were also considered but increased the run-time and showed little improvement in the mean fit error distribution. For this reason, we believe an iterative nonlinear least-squares fitting procedure like the one applied in this work is more apt for analyzing large TIPTOE datasets.

We consider a successful fit as one whose MFE is below a given error threshold. These are the shots used in the envelope analysis histograms to provide a representative picture of the measured BSV. We find that the MFE distribution for both single-burst and multi-burst results exhibit a heavy-tail towards higher MFE values indicating that some, albeit a low-percentage, do not fit well to the models used. This could be for a variety of reasons: (i) the fixed center frequency (or chirp) from the calibration step does not accurately match that of the specific shot, (ii) the amplitude of the specific shot is sufficiently low that the noise floor becomes an issue. We set the fit success threshold for our analysis to the standard metric of $\text{Median} + 1.5 \times \text{MAD}$ where MAD is the median absolute deviation of the distribution[4].

An ISO 5807 flowchart showing the logical sequence of the calibration and general iterative curve-fitting procedure which we use is shown in Fig. S3. The first iteration of the curve-fitting procedure (i.e., $n = 1$ bursts in the fit model) uses an initial guess $y_0 = \{A_1, t_{0,1}, \sigma_1\}$ consisting of the amplitude and emission time of the main peak determined by a peak-finding algorithm. The initial width is set to 15 fs. For iterations beyond the first (i.e., $n = m > 1$ bursts in the fit model), the initial guess $y_0' = \{A_1, t_{0,1}, \sigma_1, A_2, t_{0,2}, \sigma_2, \dots A_m, t_{0,m}, \sigma_m\}$ consists of the fit parameters found in the previous iteration (describing the first to $(m-1)$-th burst), as well as the parameters estimated from the envelope of the residual – the difference in envelope of the BSV field and the fit from the previous iteration. In other words, to obtain an initial guess for the amplitude and emission time of the $m$-th mode in the $m$-th iteration, the residual is calculated and fed to a simple peak-finding algorithm. The initial width is, again, 15 fs.

Single-burst and multi-burst BSV: envelope analysis

We ran our 10,000-shot dataset through the nonlinear iterative curve-fitting procedure and found 5163 shots were single-burst. The distributions of the fit parameters for the single-burst shots whose MFE is below the error threshold (see previous subsection) are shown in Fig. S4. Panel a shows the distribution of MFE, while panels b-e show the distribution of the fit parameters. The absolute phase distribution shown in panel c is wrapped between [0, 2π). The phase clusters around an offset, i.e. the phase is fixed (apart from a sign that cannot be resolved with our 800 nm reference, see above), as expected for BSV.

The distribution of the fit parameters for the multi-burst shots is presented in the main text. Figure S5a shows the distribution of the MFE, panel b shows the distribution of the FWHM of adjacent bursts, while panel c shows the distribution of the absolute phase. A similar clustering is seen in the phase of the multi-burst shots.

Simulation of overlapping Gaussian bursts

We test the performance of our nonlinear iterative curve-fitting procedure against three simulated datasets. The first test is against 10,000 single-burst shots, each of which with amplitude, emission time, and absolute phase selected from uniform distributions spanning the range of physical values expected from our experiment. The width (or HWHM) of the simulated pulses is selected from a normal distribution centered at 7 fs and with a standard deviation of 1 fs. This corresponds to an intensity FWHM of approximately 10 fs which is on the order of the measured $\sigma_I$ discussed in the manuscript.

As can be seen in Fig. S6a, the iterative curve-fitting procedure finds that each simulated shot is, in fact, single-burst. In other words, it does not succumb to overfitting and assigning multiple bursts to truly single-burst signals. The mean fit error distribution for the single-burst dataset is almost entirely within the zero bin of the histogram in panel b. This low error is corroborated by the negligible normalized root-mean-squared error (NRMSE) for each fit parameter shown in panel c.

The second dataset consists of 10,000 two-burst shots whose amplitudes, widths, emission times, and absolute phases are all selected from the same distributions as in the first test. The only added condition is that the simulated bursts are separated by at least the mean width (i.e., for this simulation 7 fs). From panel a, we see that the iterative curve-fitting procedure correctly labels 90% of the shots as two-burst while 10% are identified as single-burst. In other words, the curve-fitting algorithm found that, for 10% of the shots, adding an additional burst to the fit model did not drop the MFE by more than 0.5% and/or it was not able to add an additional burst without the two bursts being within one HWHM of one another. For the 10% of the dataset that are identified as single-burst shots, the ratio of the maximum amplitude $\max(A_1, A_2)$ to the minimum amplitude $\min(A_1, A_2)$ is, on average, $\sim 11$. This indicates that one burst in the simulation dominates the overall shot which may explain why the curve-fitting algorithm fails to identify both bursts. The amplitude distribution of BSV is not uniform so large amplitude ratios as described here should be far less common (see Fig. 2d of the main text). As depicted in panel b, the addition of a second potentially overlapping burst in the simulation results in an increase in the width of the MFE distribution. Importantly, however, this distribution is still limited to approximately 1.5% thus indicating a high-level fit performance overall.

For the shots that were correctly labelled as two-burst, we see an increase in the NRMSE for each fit parameter over the single-burst case. The NRMSE for the emission time and absolute phase are each below 0.3% indicating that the iterative curve-fitting procedure, even in the two-burst case, can effectively extract these fit parameters. The highest NRMSE value is for the width (3.48%) of each burst.

Finally, for the last test, we simulate $10{,}000$ three-burst shots, again, whose amplitudes, widths, emission times, and absolute phases are all selected from the same distributions as in the other two tests. Unsurprisingly, we find a reduction in the overall performance of the curve-fitting procedure. The number of correctly labelled shots reduces from 90% in the two-burst test, to 71% in the three-burst test. An increase in the width of the MFE distribution and a marginal increase in the NRMSE for each fit parameter are found in those shots which are correctly labelled.

Importantly, the only physically motivated feature of these simulations is that the width distribution is not uniform across a range of values but rather favors a mean value (in this case 7 fs). The distributions sampled for all other fit parameters are uniform. This implies that the unique features found in our envelope analysis of the measured BSV (e.g., the exponential, heavy-tail distribution of amplitudes, the clustering of the phase, etc.) are due to the temporal structure of the BSV itself, rather than methods used to analyze it.

Single-mode BSV photon statistics

The probability of detecting $2m$ photons in a squeezed vacuum field is described by the well-known probability distribution[5],

$$P(2m) = \frac{(2m)!}{2^{2m}(m!)^2} \frac{(\tanh r)^{2m}}{\cosh r}, \tag{3}$$

and the probability of detecting $2m + 1$ photons is,

$$P(2m+1) = 0. \tag{4}$$

The exact form of this distribution is useful for describing the photon statistics of weakly squeezed fields with low mean photon numbers, but for truly bright fields it is computationally challenging to apply due to overflow limits and stability in the leading factorial term. Instead, we use an approximation of the probability envelope to study the statistics of brightly squeezed vacuum.

We begin by recognizing the definition of the binomial coefficient,

$$\binom{n}{k} = \frac{n!}{k!\,(n-k)!}, \tag{5}$$

where we let $n = 2m$ and $k = m$ to find,

$$\binom{2m}{m} = \frac{(2m)!}{(m!)^2}. \tag{6}$$

The even number probability distribution $P(2m)$ can be written in terms of the binomial coefficient,

$$P(2m) = \frac{1}{2^{2m}} \binom{2m}{m} \frac{(\tanh r)^{2m}}{\cosh r}, \tag{7}$$

whose asymptotic form can be found by applying Stirling's approximation,

$$\begin{aligned} P(2m) &\rightarrow \frac{1}{2^{2m}} \left( \frac{2^{2m}}{\sqrt{\pi m}} \right) \frac{(\tanh r)^{2m}}{\cosh r} \\ &= \frac{1}{\sqrt{\pi m}} \frac{(\tanh r)^{2m}}{\cosh r}. \end{aligned} \tag{8}$$

The gain parameter $r$ is defined with respect to the mean photon number $\langle n \rangle$,

$$\langle n \rangle = \sinh^2(r)\,, \tag{9}$$

such that $\cosh^2(r) = \langle n \rangle + 1 \approx \langle n \rangle$ or $\cosh r \approx \sqrt{\langle n \rangle}$, for large $n$. The hyperbolic tangent in the numerator can be approximated in a similar manner,

$$\begin{aligned} \tanh r &= \sqrt{1 - \frac{1}{\cosh^2(r)}} \\ &\approx \sqrt{1 - \frac{1}{\langle n \rangle}}. \end{aligned} \tag{10}$$

Raising the above to the power of $2m$ gives,

$$\begin{aligned} (\tanh r)^{2m} &= \left(1 - \frac{1}{\langle n \rangle}\right)^m \\ &= \exp\left[\ln\left(\left(1 - \frac{1}{\langle n \rangle}\right)^m\right)\right] \\ &= \exp\left[m \ln\left(1 - \frac{1}{\langle n \rangle}\right)\right] \\ &\approx \exp\left(-\frac{m}{\langle n \rangle}\right), \end{aligned} \tag{11}$$

where the natural logarithm has been approximated by its first-order Taylor series around $1/\langle n \rangle = 0$. Substituting into the overall probability distribution, we find,

$$P(2m) = \frac{1}{\sqrt{\pi m \langle n \rangle}} \exp\left(-\frac{m}{\langle n \rangle}\right), \tag{12}$$

or in terms of $n$,

$$P(n) = \sqrt{\frac{2}{\pi n \langle n \rangle}} \exp\left(-\frac{n}{2\langle n \rangle}\right). \tag{13}$$

The probability envelope in the limit of large photon numbers is then given by $P_{\text{envelope}} = 0.5 \times \big(P(n = \text{even}) + P(n = \text{odd})\big) = 0.5 \times P(n)$ as adjacent even and odd bins average to half the value of the even bin. Using this definition, we find the form commonly used to describe BSV statistics[6],

$$P_{\text{envelope}} = \frac{1}{\sqrt{2\pi n \langle n \rangle}} \exp\left(-\frac{n}{2\langle n \rangle}\right). \tag{14}$$

In Fig. 2f, we plot this envelope with $\langle n \rangle = 0.8 \times 10^{10}$ photons determined from the separately measured average BSV shot energy of 1.5 nJ, which converts to an average BSV burst energy of 1 nJ (since there are ~1.5 bursts per shot on average). The normalized fluctuations measured by TIPTOE are linearly proportional to the true fluctuations of the perturbative field. Therefore, we scale the measured BSV burst energy distribution to the experimental value of 1 nJ and then convert it to a photon number scale to get the distribution shown in Fig. 2f.

Quantum coherence functions and the overlap of BSV bursts

The decrease of $g^{(2)}(t_1, t_2)$ from the expected value of 3 for single-mode BSV to 2 in the same-time (i.e., $t_1 = t_2$) wings of Fig. 4c can be ascribed to the overlap of, on average, two independent bursts. To probe this effect directly, we scan a window of varying width along the same-time diagonal of the second-order coherence function and calculate both the burst-averaged $g_{i,\mathrm{w}}^{(2)}(0)$ and the shot-averaged $g_{\mathrm{BSV,w}}^{(2)}(0)$ over the finite duration. This is in contrast to the reported values for $g_i^{(2)}(0)$ and $g_{\mathrm{BSV}}^{(2)}(0)$ which consider the entire temporal range.

The effect of the window in this sample calculation is two-fold. First, only those bursts whose emission time is within the finite temporal range of the window are included in the calculation. Equivalently, the wings of neighboring bursts do not affect the windowed coherence value unless the emission time of the that burst too falls within the window. In this way, the window width tunes the amount of overlap present at a given time. Second, the integration is over the burst(s) intensity envelope(s) as given by the burst envelope analysis and is performed only over the range of the window. In the shot-averaged calculation, multiple bursts in a single shot means the sum of the integral over each burst is taken and thus the effect of their overlap plays a role. Each shot contributes a single value to the mean in the overall windowed $g_{\mathrm{BSV,w}}^{(2)}(0)$. However, in the burst-averaged calculation, multiple bursts in a single shot means the integral over each burst contributes a value separately to the mean in the overall windowed $g_{i,\mathrm{w}}^{(2)}(0)$.

In Fig. S7, a and b, we find that for a window width less than or equal to $\sim 30$ fs, the burst-averaged $g_{i,\mathrm{w}}^{(2)}(0)$ and the shot-averaged $g_{\mathrm{BSV,w}}^{(2)}(0)$ agree. This width corresponds to the average separation of BSV bursts found from the envelope analysis. Therefore, for windows shorter than $\sim 30$ fs, each shot contains on average a single burst, so the two calculations are equivalent. However, beyond this width the shot-averaged $g^{(2)}_{\mathrm{BSV,w}}(0)$ decreases as the wings of neighboring bursts begin to seep into the window, the burst overlap increases on average, and the overall degree of coherence decreases. In contrast, the burst-averaged $g^{(2)}_{i,\mathrm{w}}(0)$ increases as more individual bursts contribute their stochastic fluctuations to the final mean – this can be thought of as filling in the characteristic heavy-tail distribution.

For window widths greater than $\sim 50$ fs, the two coherence values begin to converge towards different values as this is approximately the sum of the average separation of BSV bursts and the average burst intensity envelope width. At these ranges, the burst-averaged $g^{(2)}_{i,\mathrm{w}}(0)$ converges towards 3 indicative of stochastic, mode-like behavior. In contrast, the shot-averaged $g^{(2)}_{\mathrm{BSV,w}}(0)$ converges towards slightly above 2 because each shot contains on average 1.5 bursts. This agrees with $g^{(2)}_{\mathrm{BSV}}(0)$ which uses the full envelope of the measured BSV field.

Whittaker-Shannon decomposition of two-crystal BSV

*Brief background.* Recently, the work presented in ref.[7] applied the methods of the Whittaker-Shannon interpolation to BSV and found that it provides a localized description of the temporal structure compared to the more commonly used Schmidt mode decomposition[5]. The general idea is as follows: first, assume the squeezed state takes the following form,

$$|\Psi\rangle = \exp\left[\frac{\beta}{2}\int dt_1 dt_2 \bar{\gamma}(t_1,t_2)\bar{a}^{\dagger}(t_1)\bar{a}^{\dagger}(t_2) - \mathrm{H.c.}\right]|v\rangle, \tag{15}$$

where $\bar{\gamma}(t_1,t_2)$ is the temporal analogue of the more commonly used Joint Spectral Amplitude (JSA) or $\gamma(\omega_1,\omega_2)$, and $|v\rangle$ denotes the vacuum state. The two-dimensional function $\bar{\gamma}(t_1,t_2)$ is called the Joint Temporal Amplitude (JTA) and is related to the JSA via a two-dimensional Fourier transformation. The parameter $|\beta|$ controls the squeezing strength.

If the JSA is approximately band-limited, it is essentially nonzero outside of the range $\omega_1,\omega_2 \in [-\Omega/2,\Omega/2]$, where $\Omega > 0$, then the JTA can be decomposed according to the Whittaker-Shannon sampling theorem. That is, a time-dependent function $\bar{f}(t)$ can be approximated by the sum of its value at specific times $n\tau$, where $\tau = 2\pi/\Omega$, and with sinc-dependent weights,

$$\bar{f}(t) = \sum_n \bar{f}(n\tau)\mathrm{sinc}\left(\frac{(t-n\tau)\pi}{\tau}\right). \tag{16}$$

The Whittaker-Shannon modes are then defined as the following,

$$\bar{\chi}_n(t) = \frac{1}{\sqrt{\tau}}\mathrm{sinc}\left(\frac{\pi(t-n\tau)}{\tau}\right). \tag{17}$$

Applying the Whittaker-Shannon sampling theorem to the JTA gives the following representation,

$$\bar{\gamma}_{\mathrm{WS}}(t_1,t_2) \approx \tau\sum_{n,m}\bar{\gamma}(n\tau,m\tau)\,\bar{\chi}_n(t_1)\bar{\chi}_m(t_2), \tag{18}$$

which is the decomposition of the JTA in terms of the Whittaker-Shannon modes.

The two main results of ref.[7] which we are interested in here are, first, the decomposition of the first- and second-order time-dependent (non-normalized) correlation functions given by Eq. 6.6 and 6.7 in ref.[7], respectively. To write these explicitly,

$$\bar{G}^{(1)}(t_1,t_2) = \bar{\boldsymbol{\chi}}^{\dagger}(t_1)\sinh^2(\boldsymbol{P})\,\bar{\boldsymbol{\chi}}(t_2), \tag{19}$$

and

$$\begin{aligned}\bar{G}^{(2)}(t_1,t_2) = {} & |\bar{\boldsymbol{\chi}}^{T}(t_1)\boldsymbol{U}\sinh(\boldsymbol{P})\cosh(\boldsymbol{P})\,\bar{\boldsymbol{\chi}}(t_2)|^2 \\ & +\bar{G}^{(1)}(t_1,t_1)\bar{G}^{(1)}(t_2,t_2) + \left|\bar{G}^{(1)}(t_1,t_2)\right|^2.\end{aligned} \tag{20}$$

This is where the set of displaced Whittaker-Shannon modes $\bar{\chi}_n(t)$ forms the column vector $\bar{\boldsymbol{\chi}}(t)$ at a specific time $t$. In this formalism, $\boldsymbol{P}$ and $\boldsymbol{U}$ are related by the left polar decomposition of $\boldsymbol{\beta}$ where $\beta_{nm} = \beta\tau\bar{\gamma}(n\tau,m\tau)$ (see ref.[7] or ref.[8] for more information).

Second, the authors find a packet expansion of the same-time first-order correlation function $\bar{G}^{(1)}(t_1=t_2)$. The decomposition is as follows,

$$\bar{G}^{(1)}(t_1=t_2) = \sum_n \Gamma_n^2|\bar{\rho}_n(t)|^2\,, \tag{21}$$

which is particularly useful as $\bar{G}^{(1)}(t_1 = t_2)$ represents the photon density of the pulse. This result is analogous to the expansion of the photon density in terms of Schmidt modes. With respect to the width of the pump, however, the functions $\{\bar{\rho}_n(t)\}$ remain relatively localized in comparison to the well-known Schmidt modes which become highly nonlocal as the order increases (for an in-depth comparison, see ref.[7]).

*Derivation of the JSA for the two-crystal setup.* We begin by examining the case of a single crystal before generalizing the result to include two crystals separated by an air gap. Below, we follow the general procedure provided in ref.[9] for deriving the JSA of BSV in a nonlinear interferometer.

We consider a classical Gaussian pump with field $E_{0,p} \exp\left(-t^2/2\tau_p{}^2\right) e^{i(k_p r - \omega_p t)}$. Under plane-wave expansion of the signal and idler fields, the Hamiltonian describing PDC generation in a $\chi^{(2)}$ crystal is given by the following,

$$H \propto i \iiint d\boldsymbol{k}_s d\boldsymbol{k}_i d^3\boldsymbol{r} \chi^{(2)}(\boldsymbol{r}) E_{0,p} e^{-\frac{t^2}{2\tau_p{}^2}} A_{\boldsymbol{k}_s} A_{\boldsymbol{k}_i} e^{i(\Delta\boldsymbol{k}\cdot\boldsymbol{r} - \Delta\omega t)} a_{\boldsymbol{k}_s}^\dagger a_{\boldsymbol{k}_i}^\dagger + \text{H.c.} \tag{22}$$

where $\Delta\boldsymbol{k} = \boldsymbol{k}_p - \boldsymbol{k}_s - \boldsymbol{k}_i$ and $\Delta\omega = \omega_p - \omega_s - \omega_i$ *(49)*. In the collinear propagation approximation, the transverse components of the wavevectors in the Hamiltonian can be neglected. For the pump wavevector, for example, this approximation implies $|\boldsymbol{k}_p \cdot \boldsymbol{r}| \approx k_{p_z} z$ where $k_{p_z} = n(\omega_p)\omega_p/c$ (i.e., is a function dependent on $\omega_p$). We drop the $z$ on the longitudinal components for notational simplicity. As such, the highly nested integral in the Hamiltonian becomes an integral over angular frequency and $z$. Following the work presented in ref.[9], we make the zero-energy mismatch approximation such that the pump pulse with bandwidth $\Omega_p = 1/\tau_p$ takes the following form in frequency space,

$$\begin{aligned} FT\{E_p\} &= E_{0,p} e^{ik_p z} \int d\omega \delta(\omega - \omega_s - \omega_i) e^{-\frac{(\omega-\omega_p)^2}{2\Omega_p{}^2}} e^{-i\omega t} \\ &= E_{0,p} e^{ik_p z} e^{-\frac{(\Delta\omega)^2}{2\Omega_p{}^2}} e^{-i(\omega_s+\omega_i)t}, \end{aligned} \tag{23}$$

which is now a function of the signal and idler frequencies. We assume $\chi^{(2)}(\boldsymbol{r})$ varies only along the depth of the nonlinear medium z. Substituting this into the Hamiltonian, we arrive at the final form,

$$H \propto iE_{0,p} \iiint d\omega_s d\omega_i dz e^{-\frac{(\Delta\omega)^2}{2\Omega_p{}^2}} \chi^{(2)}(z) e^{i\Delta kz} a^\dagger_{\omega_s} a^\dagger_{\omega_i} + \text{H.c.} \tag{24}$$

For a single crystal, we have $\chi^{(2)}(z) = \chi_1^{(2)}$ for $z \in [-L, 0]$. The integral over $z$ is then solved straightforwardly,

$$\begin{aligned} \int dz \chi^{(2)}(z) e^{i\Delta kz} &= \chi_1^{(2)} \int_{-L}^{0} dz e^{i\Delta kz} \\ &= \chi_1^{(2)} \frac{1}{i\Delta k}\left(1 - e^{-i\Delta kL}\right) \\ &= \chi_1^{(2)} L\text{sinc}\left(\frac{\Delta kL}{2}\right) e^{-i\frac{\Delta kL}{2}}, \end{aligned} \tag{25}$$

and thus, the JSA is proportional to,

$$\gamma(\omega_s, \omega_i) \propto \text{sinc}\left(\frac{\Delta kL}{2}\right) e^{-\frac{(\Delta\omega)^2}{2\Omega_p{}^2}} e^{-i\frac{\Delta kL}{2}}. \tag{26}$$

Under these approximations, we can generalize the single crystal solution to the two-crystal solution by treating the effect of the air gap as a mere additional phase shift due to the longitudinal wavevector mismatch over the thickness $L_{\text{air}}$[10]. Since $\chi^{(2)}$ for air is negligible, the integral over all $z$ is replaced by the sum of two integrals, one for each crystal, with the second integral picking up the additional phase mismatch factor,

$$\begin{aligned} \int dz \chi^{(2)}(z) e^{i\Delta kz} \rightarrow \chi_1^{(2)} \int_{-L}^{0} dz e^{i\Delta kz} + \chi_2^{(2)} e^{i\Delta k_{\text{air}} L_{\text{air}}} \int_{0}^{L} dz e^{i\Delta kz} \\ = \chi_1^{(2)} L\text{sinc}\left(\frac{\Delta kL}{2}\right) e^{-i\frac{\Delta kL}{2}} \\ + \chi_2^{(2)} e^{i\Delta k_{\text{air}} L_{\text{air}}} L\text{sinc}\left(\frac{\Delta kL}{2}\right) e^{i\frac{\Delta kL}{2}} \end{aligned} \tag{27}$$

Assuming the two crystals are of the same uniaxial material, with the second having its crystal axis flipped with respect to the first, then $\chi_2^{(2)} = -\chi_1^{(2)}$ and the final expression simplifies to:

$$\int dz \chi^{(2)}(z) e^{i\Delta kz} \rightarrow Lsinc\left(\frac{\Delta kL}{2}\right)\left(\chi_1^{(2)} e^{-i\frac{\Delta kL}{2}} + \chi_2^{(2)} e^{i\Delta k_{\text{air}} L_{\text{air}}} e^{i\frac{\Delta kL}{2}}\right)$$

$$= \chi_1^{(2)} L sinc\left(\frac{\Delta kL}{2}\right)\left(e^{i\left(\frac{\Delta kL+\Delta k_{\text{air}}L_{\text{air}}}{2}\right)} - e^{-i\left(\frac{\Delta kL+\Delta k_{\text{air}}L_{\text{air}}}{2}\right)}\right) e^{-i\Delta kL} e^{-i\left(\frac{\Delta k_{\text{air}}L_{\text{air}}}{2}\right)}$$
$$= 2i\chi_1^{(2)} L sinc\left(\frac{\Delta kL}{2}\right) e^{-i\left(\frac{\Delta kL}{2}\right)}$$
$$\times \sin\left(\frac{\Delta kL + \Delta k_{\text{air}}L_{\text{air}}}{2}\right) e^{-i\left(\frac{\Delta kL+\Delta k_{\text{air}}L_{\text{air}}}{2}\right)}. \quad (28)$$

The JSA for this setup is then proportional to,

$$\gamma(\omega_s, \omega_i) \propto sinc\left(\frac{\Delta kL}{2}\right) e^{-\frac{(\Delta\omega)^2}{2\Omega_p{}^2}} \sin\left(\frac{\Delta kL + \Delta k_{\text{air}}L_{\text{air}}}{2}\right) e^{-i\left(\frac{\Delta kL+\Delta k_{\text{air}}L_{\text{air}}}{2}\right)} e^{-i\frac{\Delta kL}{2}}. \quad (29)$$

For our setup, we have two crystals of length $L = 2$ mm separated by an air gap of length $L_{\text{air}} \approx 19$ cm. The pump width is set to 50 fs in the calculations.

*Estimating $\bar{g}^{(1)}(t_1, t_2)$ and $\bar{g}^{(2)}(t_1, t_2)$ for two-crystal BSV.* Using the JSA found in the previous subsection, we can transform it to the time-domain to define the approximate JTA (and the Joint Temporal Intensity, JTI) of our setup. We set the bandlimit to the range around the degenerate frequency where the modulus squared of the JSA (the Joint Spectral Intensity, JSI) is greater than 1% of its maximum value, such that we capture the non-negligible frequency content of the JSA. The value of the bandlimit is $\sim 0.81$ rad/fs or 0.13 PHz. The Whittaker-Shannon mode spacing and duration (FWHM) are then approximately 7.8 fs and 9.4 fs, respectively.

To apply the results of ref.[7], we need to set the squeezing parameter $\beta$ to ensure we are in the same squeezing regime as our experiment. Given that our BSV contains on the order of $10^{10}$ photons per pulse, we tune $\beta$ until the integral over the photon density $\bar{G}^{(1)}(t_1 = t_2)$ is $\sim 1 \times 10^{10}$. This is equivalent to tuning $\beta$ until the sum over the weighting factors $\sum_n \Gamma_n^2$ is equal to $\sim 1 \times 10^{10}$. The precise $\beta$ used was 24.15.

We use the extended Sellmeier equations for BBO to calculate the wavelength-dependent refractive index in the JSA[11]. In addition, we define the time-zero using the reference frame of the BSV rather than the pump by shifting the JTA by the theoretical relative group delay,

$$\Delta t_{\text{g}} = \frac{L}{c}\Delta n_{\text{g}}. \quad (30)$$

We perform this shift in the frequency domain by multiplying the derived JSA by an additional linear phase with slope $\Delta t_{\mathrm{g}}$.

Throughout this analysis, we use normalized JSA and JTI distributions. Figure S8, a and b, show $|\gamma(\omega_s, \omega_i)|^2$ and $|\bar{\gamma}(t_1, t_2)|^2$, respectively. The values sampled according to the Whittaker-Shannon sampling theorem are shown in c while the reconstruction using these sampled values according to Equation 18 is given in d. Using the expansion of the same-time first-order coherence function we decompose the photon density into constituent packet expansions as described in ref.[7] This is shown in Fig. S9a along with the contributions from the five most dominant packets as defined by the weighting factors $\Gamma_n^2$. The distribution of $\Gamma_n^2$ is plotted in the bar chart of b. We estimate the width of each packet whose $\Gamma_n^2$ is above 10% of the maximum weighting factor (i.e., $\Gamma_0^2$) by fitting it to a Gaussian function. We find the mean packet width to be $(15.77 \pm 0.05)$ fs which agrees with the $\sim 17$ fs width of the BSV bursts measured experimentally. In addition, the general behavior of both decreasing packet amplitude and increasing packet overlap near the wings of the photon density curve are corroborated by the experimental data. Importantly, these results do not show shot-to-shot statistics but are rather a decomposition of the state.

We can estimate the normalized first- and second-order time-dependent correlation functions according to the typical definitions,

$$\bar{g}^{(1)}(t_1, t_2) = \frac{\bar{G}^{(1)}(t_1, t_2)}{\sqrt{\bar{G}^{(1)}(t_1, t_1)\bar{G}^{(1)}(t_2, t_2)}} \tag{31}$$

and

$$\bar{g}^{(2)}(t_1, t_2) = \frac{\bar{G}^{(2)}(t_1, t_2)}{\bar{G}^{(1)}(t_1, t_1)\bar{G}^{(1)}(t_2, t_2)}. \tag{32}$$

The measurement of such time-resolved correlations is shown in the main manuscript. Importantly, for comparison between the real part (imaginary part or phase) of the theoretical $\bar{g}^{(1)}(t_1, t_2)$ and the experimental first-order coherence function, it must be multiplied by the carrier phase which would have been removed by taking the Fourier transform of the JSA over the bandwidth centered at the degenerate frequency (i.e., the Fourier shift theorem). The second-order coherence function is insensitive to this additional carrier phase term.

Towards PHz quantum random number generation

*Randomness tests.* The inherent randomness in the phase of each temporal burst in BSV, in combination with the femtosecond timescale of our measurement scheme, paves the way for QRNG at unprecedented PHz rates. Here, we provide an example of how the phase-difference between adjacent temporal bursts in a BSV dataset can be binarized to produce a stream of random bits. We then verify the randomness of our bitstream by feeding it to the NIST Test Suite – a statistical package which searches for non-randomness within a supposedly random sequence using a variety of tests[12].

From the 3881 multi-burst BSV shots, we have 3273 two-burst, 528 three-burst, and 80 four-burst shots. We are interested in the phase-difference between adjacent burst pairs within these shots. Each two-burst shot provides one phase-difference value while each three-burst and four-burst shot provide two and three phase-difference values, respectively. This corresponds to $N = 4569$ individual phase-difference values which we then binarize. The phase-difference of adjacent temporal bursts is found to cluster into two main peaks, separated by $\pi$ rads (see main text for more information). We assign any phase-difference within $\pi/2$ rads of the peak near $\pi$ a bit value of 1, and any other phase-difference a bit value of 0. This results in a bitstream containing 2241 0s and 2328 1s. This corresponds to a sample proportion of $p_1 = 2328 / 4569 = 0.51$ which is within statistical error of the expected value $p_1 = 0.5$ (where the statistical error is the inverse square root of the total number of bits[13], here $\sigma_{\text{stats}} = 1/\sqrt{4569} \approx 0.01$). While this verifies that our sequence satisfies basic statistical randomness, further investigation using the NIST Test Suite is required to rule out underlying non-randomness.

Figure S10 shows the result of seven statistical tests performed by the NIST Test Suite. If the computed P-value is greater than or equal to the minimum threshold called the significance level, $\alpha$, the sequence is considered random with a confidence of $100 \times (1 - \alpha)$. The NIST Test Suite uses a significance level of $\alpha = 0.01$ meaning a sequence will only pass as “random” if the confidence is at least 99%. Here, the P-value signifies the probability that an ideal RNG would have produced a bitstream “less random” than the bitstream being evaluated assuming a certain type of non-randomness which is test-dependent (the formal definition of the P-value relates to the null hypothesis and can be found at ref.[12]).

The **Frequency (Monobit) Test** evaluates the proportion of 0s and 1s for the binary sequence, and compares it to the expected proportion for a truly random sequence of length N. The sequence must first pass this test before it can pass any subsequent test for randomness. By extension, the **Frequency Test within a block** evaluates the frequency of 1s within an M-bit block of the sequence. Assuming the sequence is random, there should be approximately M/2 1s within the block. The sequence is divided into $\mathcal{N}$ non-overlapping blocks, and any unused bits are discarded. The block size is chosen such that $M \geq 20$, $M \geq 0.01 \times N$, and $\mathcal{N} < 100$. The frequency of 1s is then evaluated within each block. To continue, the **Runs Test** counts the number of uninterrupted sequences of 0s or 1s (i.e., the number of runs) within the total sequence. It monitors the oscillation between 0s and 1s to ensure the rate is not far more or far less than what would be expected for a random sequence. The **Longest Run of Ones in a Block Test** is similar to the **Runs Test** but performed within M-bit blocks of the total sequence. Next, the **Approximate Entropy Test** evaluates the frequency of overlapping blocks with lengths $M$ and $M + 1$ and compares it to that of a random sequence. Finally, the **Cumulative Sums Test** calculates the cumulative sum of partial sequences within the total sequence and determines if it is larger or smaller than that expected from a random sequence. The detailed step-by-step procedure, the precise test statistic, and the reference distribution for each of the tests are given in the NIST documentation[12]. We do not test the randomness of our bitstream using the other tests in the NIST Test Suite which require more than N bits to function. Acquiring a dataset containing a much larger number of multi-burst BSV shots will allow one to apply these other tests.

*Reaching PHz rates.* While our current setup allows us to measure primarily single- and two-burst BSV, a longer pump pulse will allow us to measure more temporal bursts per shot. Using the binarization procedure described here, we extract one bit of information for each temporal burst pair in every BSV shot. A sequence of shots containing purely two-burst BSV then provides one bit of information at the rate of the pump system used to generate the squeezed light. However, as the number of bursts per shot increases beyond two, the rate of bit *generation* within each shot reaches the PHz scale. This is because the burst spacing is on the order of femtoseconds. If one can resolve the phase of each burst within each shot, as we have demonstrated, then one can generate random bit sequences in near-PHz *bursts* (the time between each burst of bits would still be limited by the pump rate). This will result in quantum random-number generation with a temporal resolution and rate that was previously unattainable in electronics and other photonics approaches.

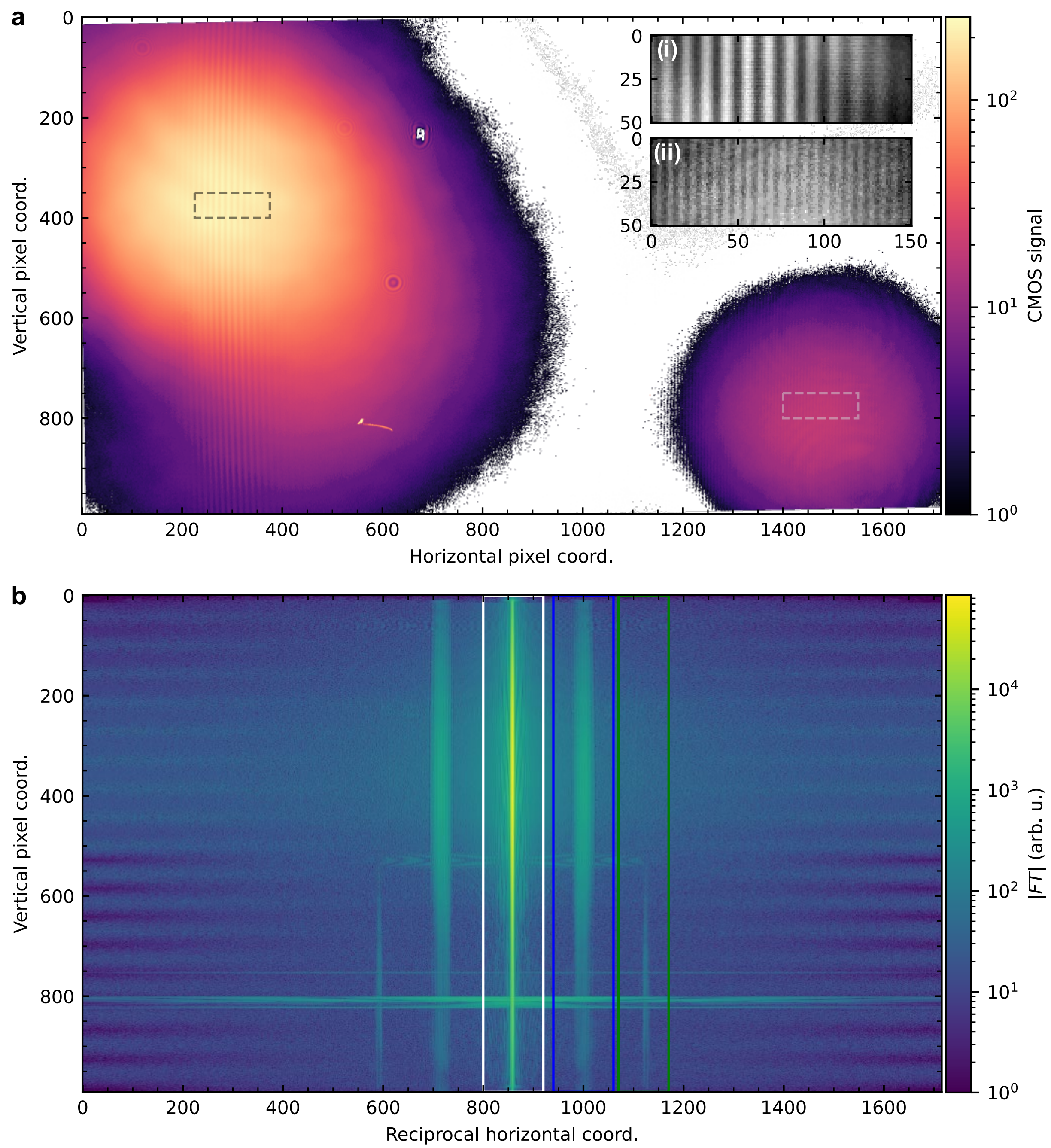

**Fig. S1: Raw signal of TIPTOE interference on CMOS sensor**

**a,** The real-space regions of interference of the two 800 nm beams (bottom-right) and two 1.6 $\mu$m beams (top-left). The inset panels (i) and (ii) correspond to zoomed-in, normalized regions of each interference pattern as marked by the black and white dashed boxes, respectively. The fringe spacing in (ii) is half than in (i), as expected. **b,** The result of a 1D Fourier transformation along the horizontal real-space axis. The white band denotes the DC offset signal while the blue and green bands denote the 1.6 μm and the 800 nm interference signals, respectively.

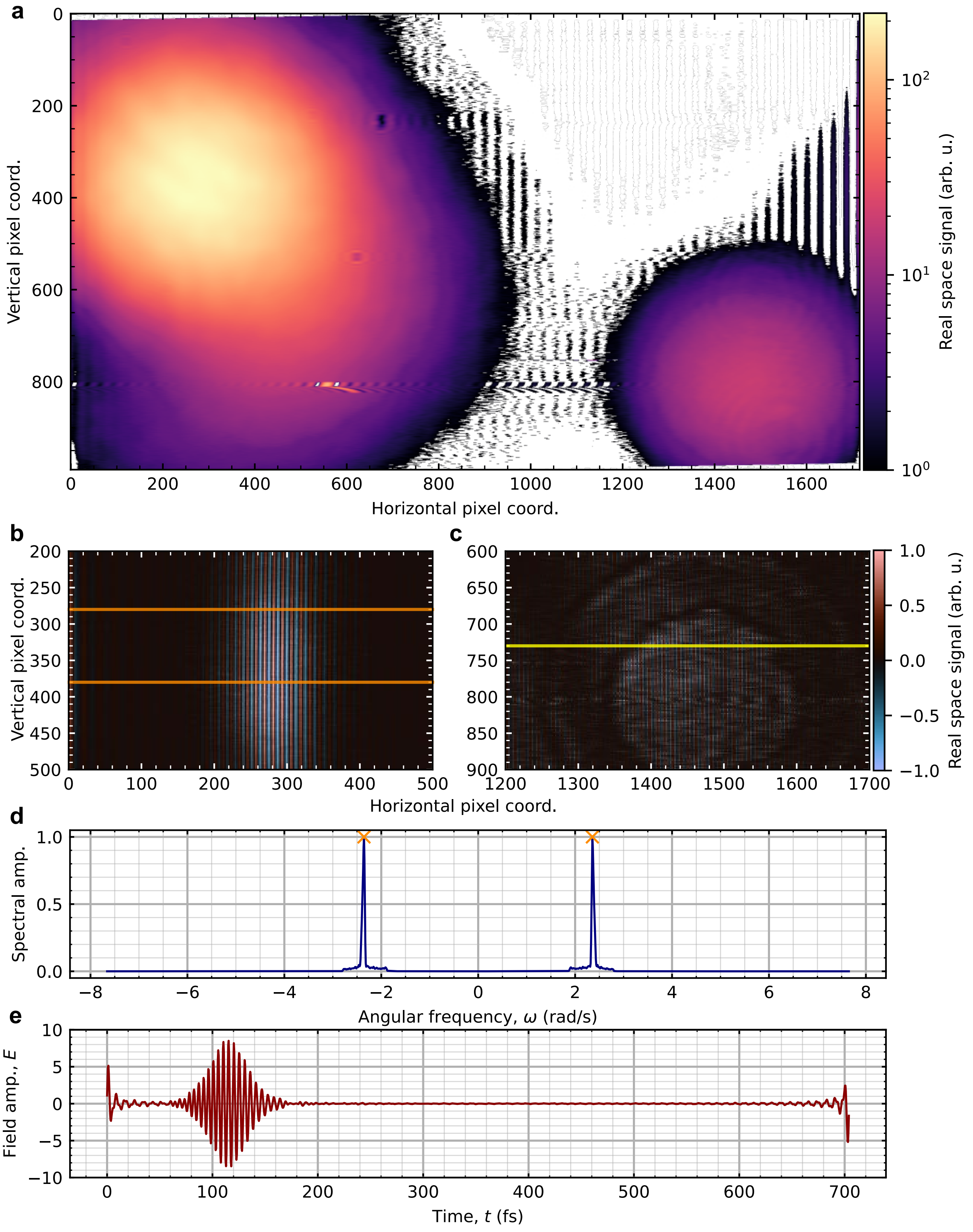

**Fig. S2: Isolation of the weak BSV signal**

The result of a 1D inverse Fourier transformation of the DC offset ROI (white region, panel **a**), of the TIPTOE interference ROI (blue region, panel **b**), and of the reference interference signal

ROI (green region, panel **C**), shown in Fig. S1**b**. In **a**, the left beam corresponds to the nonlinear absorption signal of the coherent pump, while the right beam is the overlap of the individual reference beams, but not their interference. In **b**, the real-space fringe pattern is used to directly read off the weak BSV temporal waveform by taking an average over the set of rows between the orange lines. In **c**, the real-space fringe pattern defines the reference lineout taken along the yellow line. **d,** The mean angular frequency spectrum of 100 reference signals. The spacing between the two peaks (orange ‘x’ markers) is used to calibrate the time axis (see text for more information). **e,** A sample shot of the real part of the measured BSV temporal waveform.

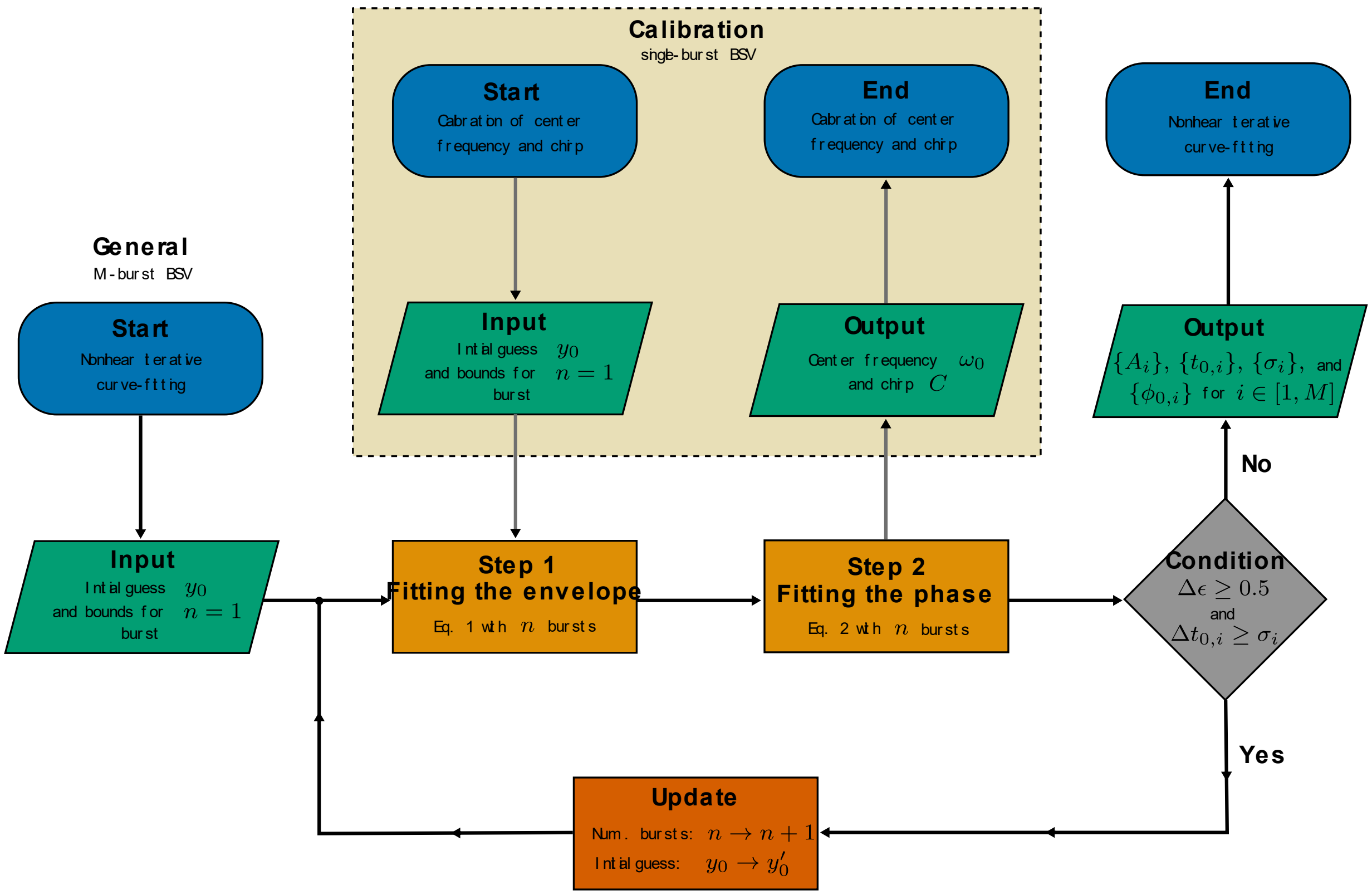

**Fig. S3: Nonlinear iterative curve-fitting flowchart**

An ISO 5807 flowchart depicting the logical sequence and decision-making steps used in the iterative curve-fitting procedure. For more information on the implementation of each step in the flowchart, see the section on curve-fitting procedure in the Supplementary Discussion.

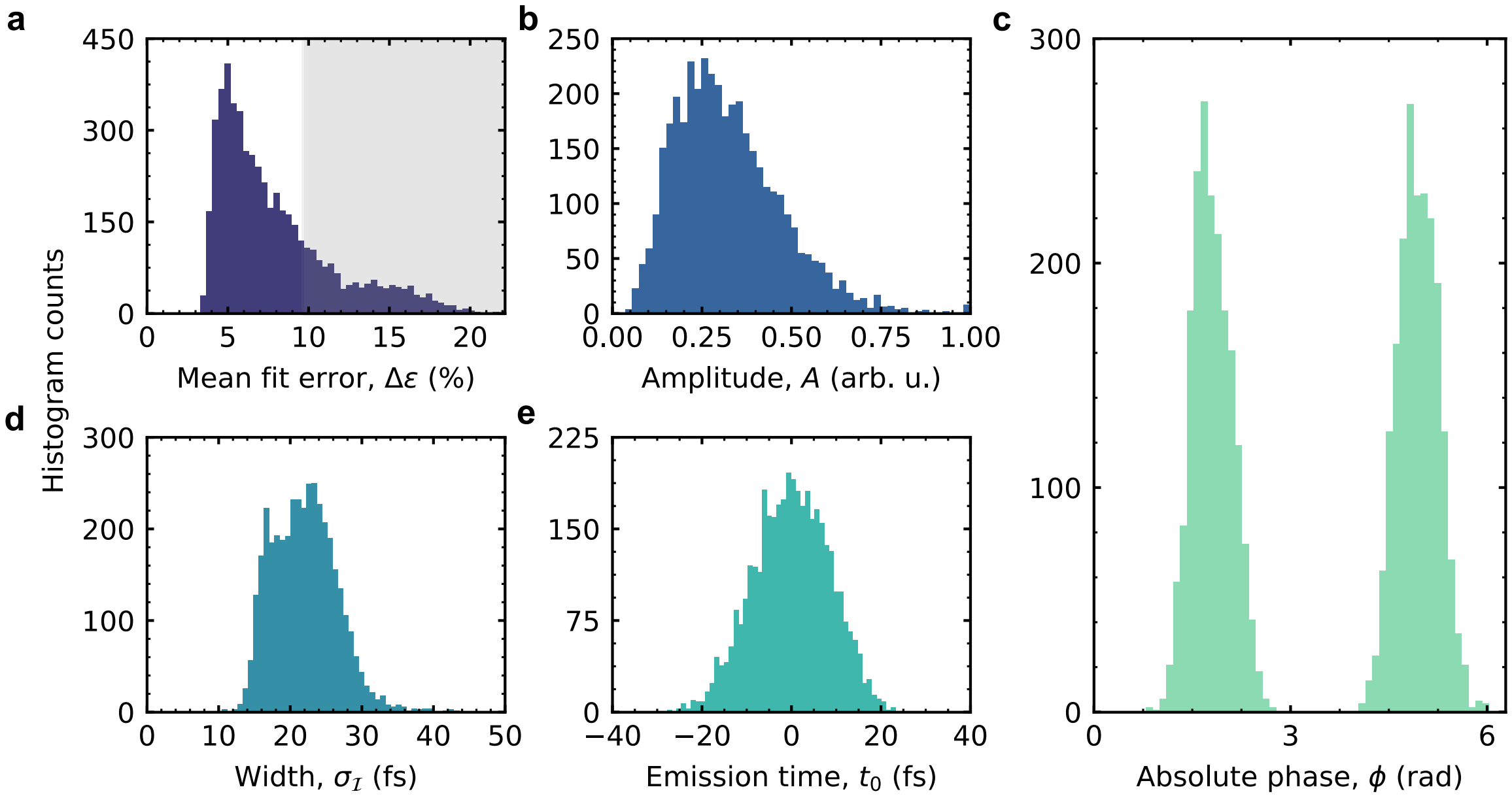

**Fig. S4: Single-burst BSV envelope analysis**

**a,** The distribution of the MFE for the single-burst shots. The grey region corresponds to shots beyond the error threshold of $\mathrm{Median} + 1.5 \times \mathrm{MAD} = 9.72\%$. **b,** The distribution of the normalized amplitude of those shots below the error threshold. **c,** The distribution of the absolute phase of those shots below the error threshold. **d,** The distribution of the width of those shots below the error threshold. **e,** The distribution of the emission time of those shots below the error threshold.

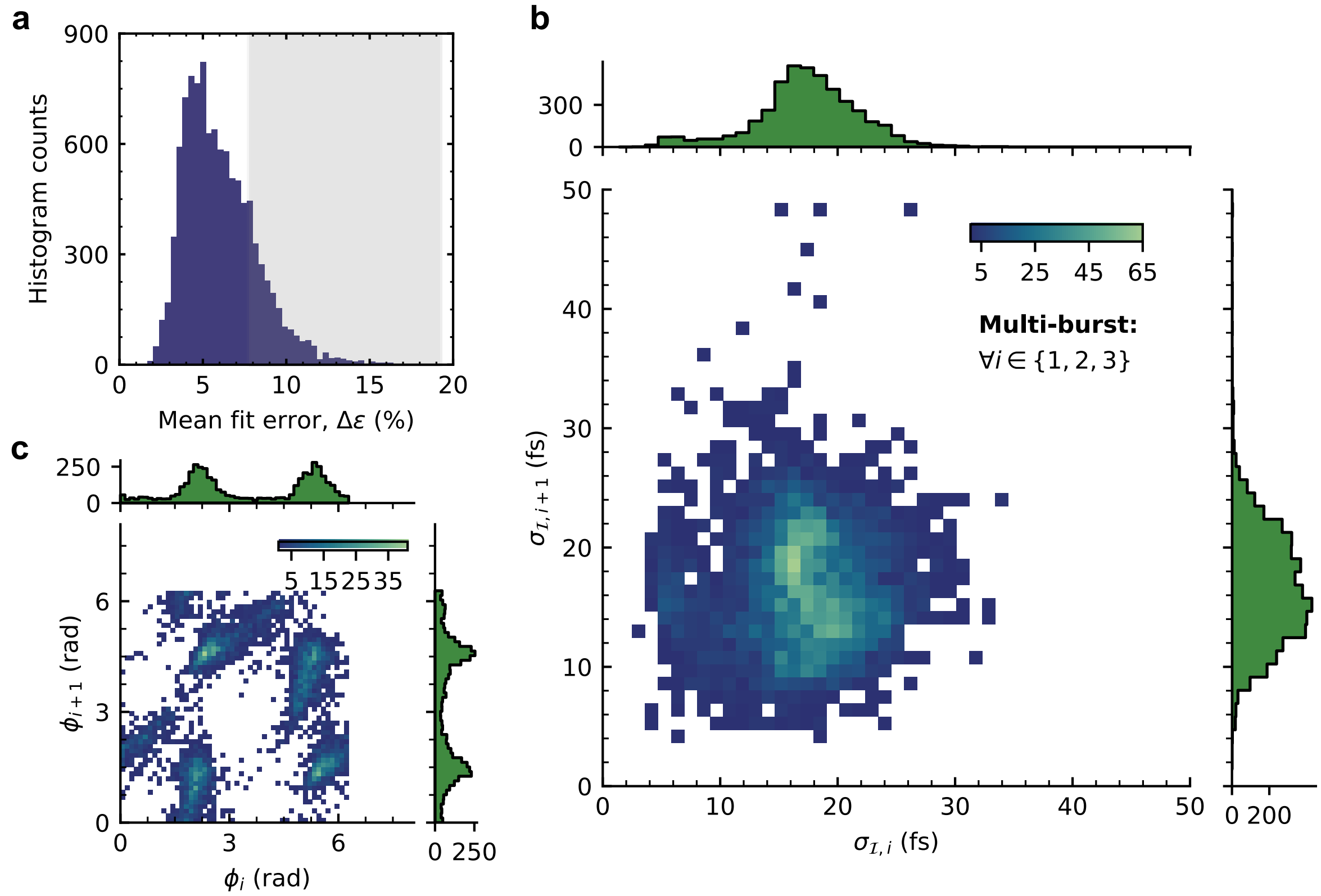

**Fig. S5: Multi-burst BSV envelope analysis**

**a,** The distribution of the MFE for the multi-burst shots. The grey region corresponds to shots beyond the error threshold of Median $+ 1.5 \times$ MAD $= 7.75\%$. **b,** The distribution of the FWHM width of the intensity envelopes of adjacent burst pairs. **c,** The distribution of the absolute phase for adjacent burst pairs. The distribution of the other fit parameters is presented in the main text. Plotting done with the seaborn package[14].

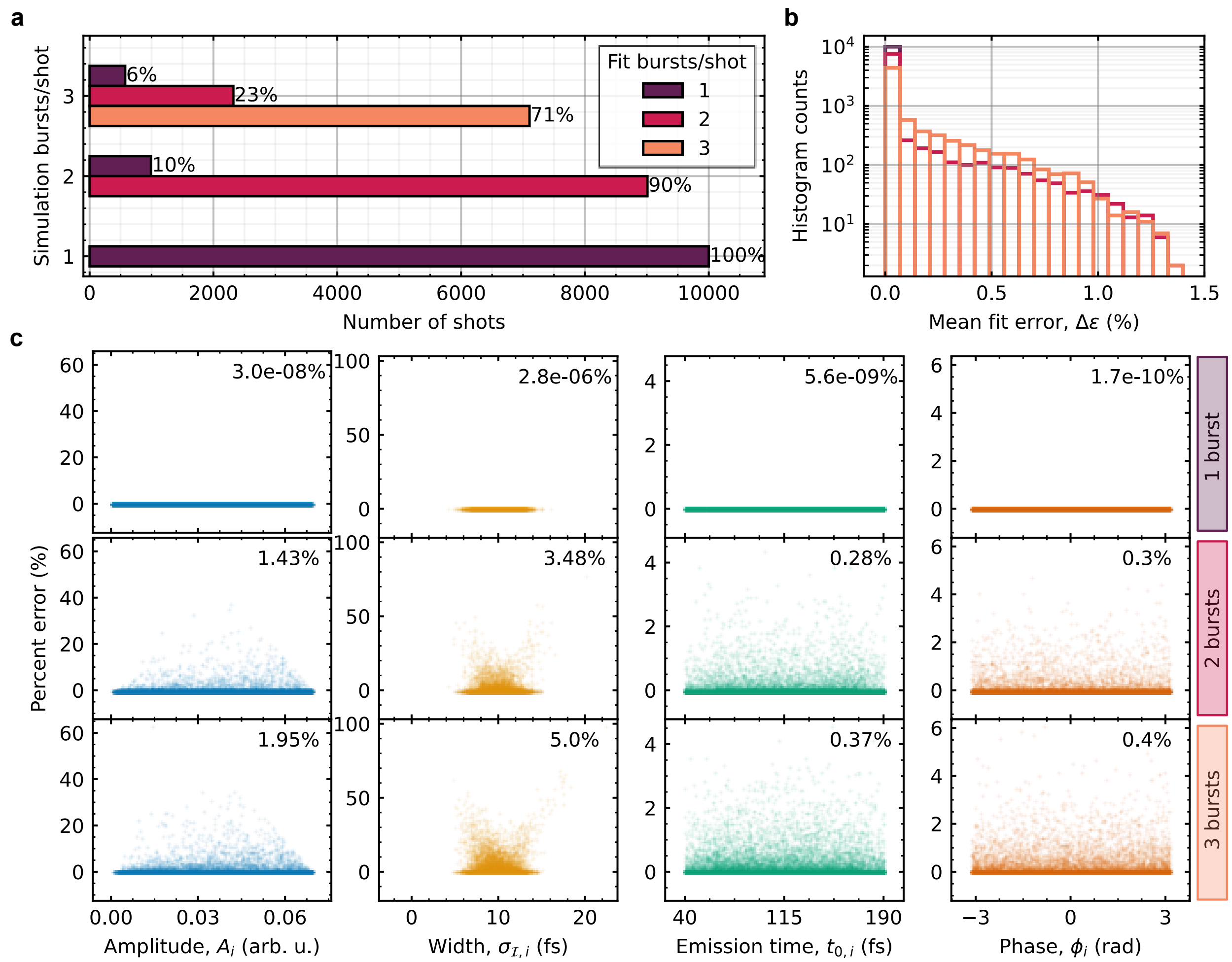

**Fig. S6: Simulation tests for evaluating the performance of the iterative curve-fitting procedure**

**a,** The distribution of the number of bursts in each shot, as estimated by feeding three different 10,000 shot simulation datasets to the iterative curve-fitting algorithm. The y-axis shows the number of simulated bursts in each dataset while the x-axis shows the number of shots found with each number of bursts, as defined by the colors in the legend. **b,** The distribution of the MFE for the three simulation datasets. The scale is logarithmic as all three sets cluster closely to the zero MFE bin, but there is noticeable change in the width of the distributions with increasing number of simulated bursts (see text for more information). **c,** The variation in percent error between the fit parameters and the true parameters as a function of the number of bursts. The number in the top-right of each subplot is the normalized root-mean-square error defined as a percentage of the range.

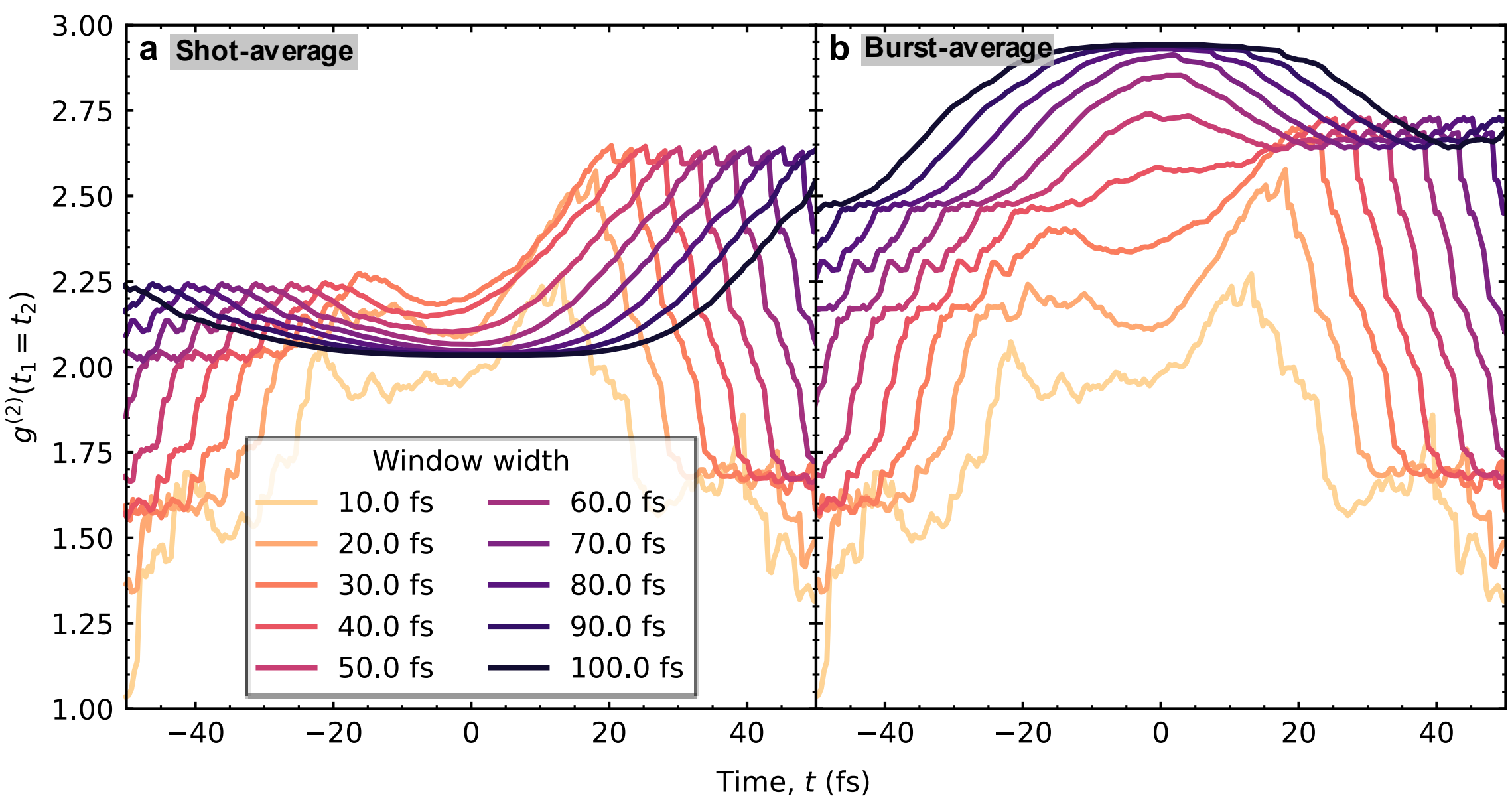

**Fig. S7: Windowed shot-average and burst-average coherence**

**a,** The windowed shot-average second-order coherence function $g^{(2)}_{\mathrm{BSV,w}}(0)$ for varying window widths. **b,** The windowed burst-average second-order coherence function $g^{(2)}_{i,w}(0)$ for varying window widths.

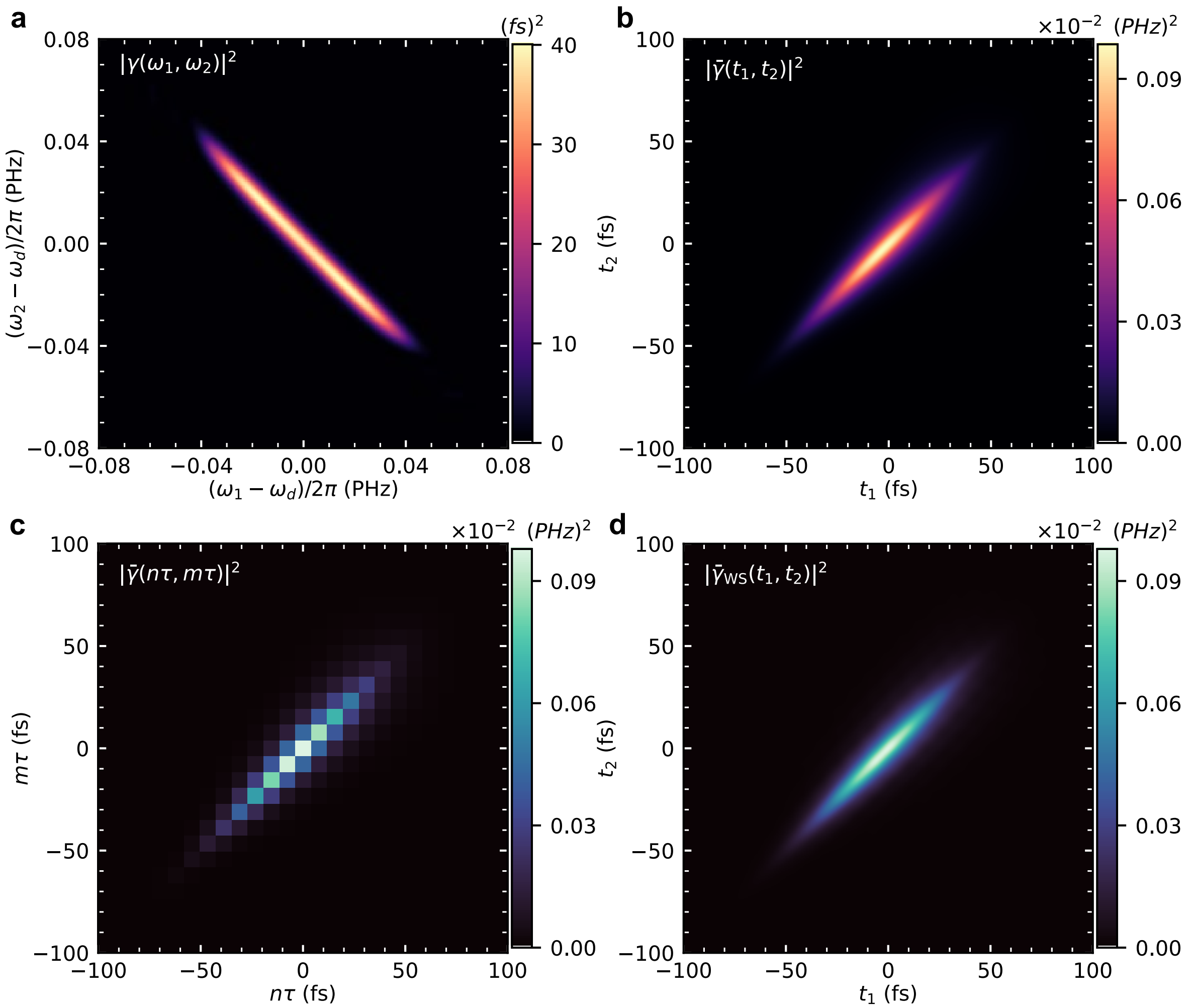

**Fig. S8: Whittaker-Shannon decomposition of two-crystal BSV**

**a,** The magnitude squared of the Joint Spectral Amplitude of our two-crystal BSV setup as described in the text. **b,** The Joint Temporal Intensity calculated from the 2D Fourier transform of the JSA. **c,** The sampled Joint Temporal Intensity. **d,** The reconstruction of the Joint Temporal Intensity using the sampled positions shown in **c** and the formula given in Equation 18.

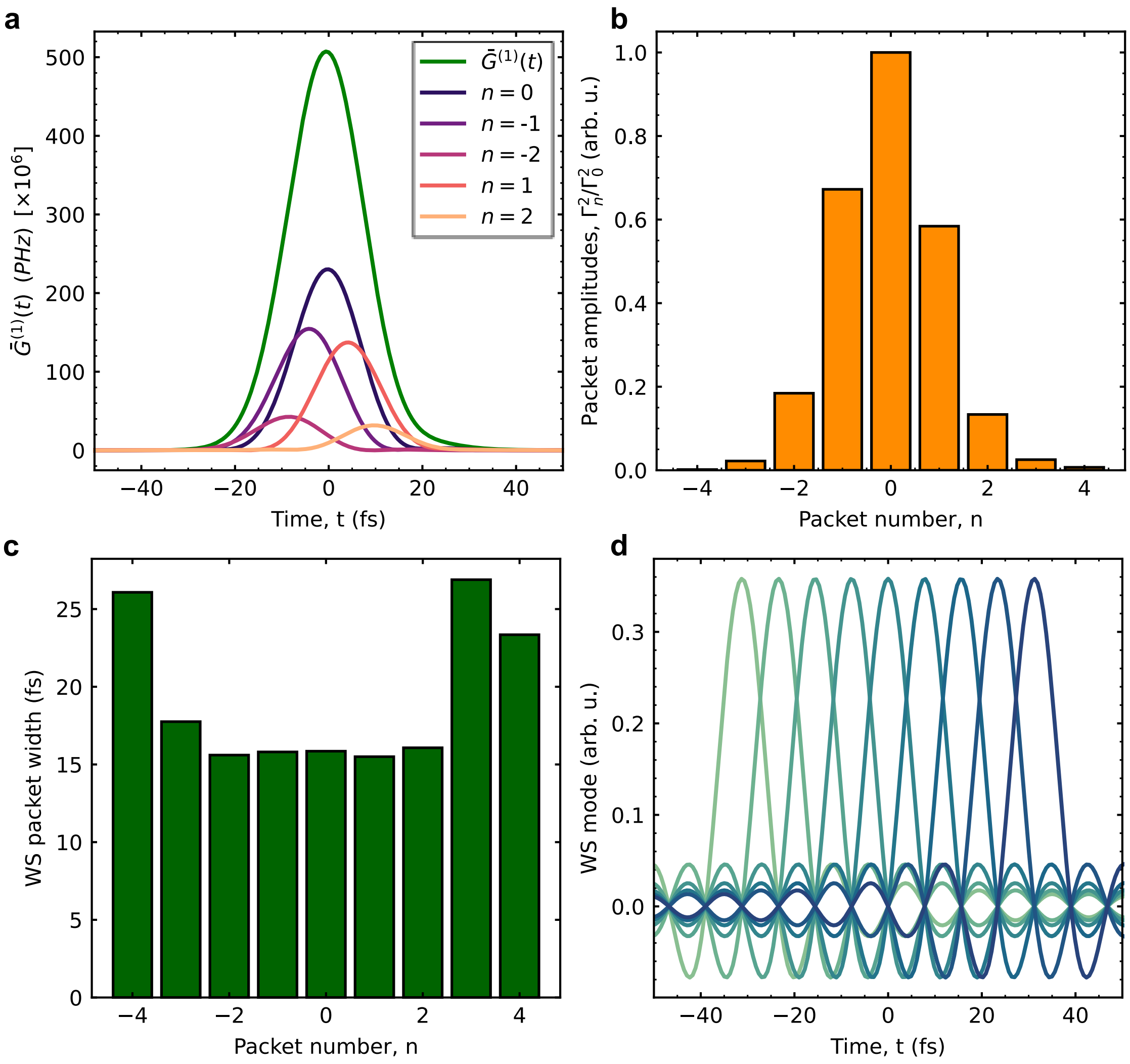

**Fig. S9: Whittaker-Shannon modes and packets**

**a,** The same-time first-order correlation function and its underlying packets. **b,** The distribution of the packet weights, normalized by the weight of the $n = 0$ packet. **c,** The distribution of the approximate packet widths. **d,** A sample of Whittaker-Shannon modes demonstrating their spacing of $\tau$ and widths.

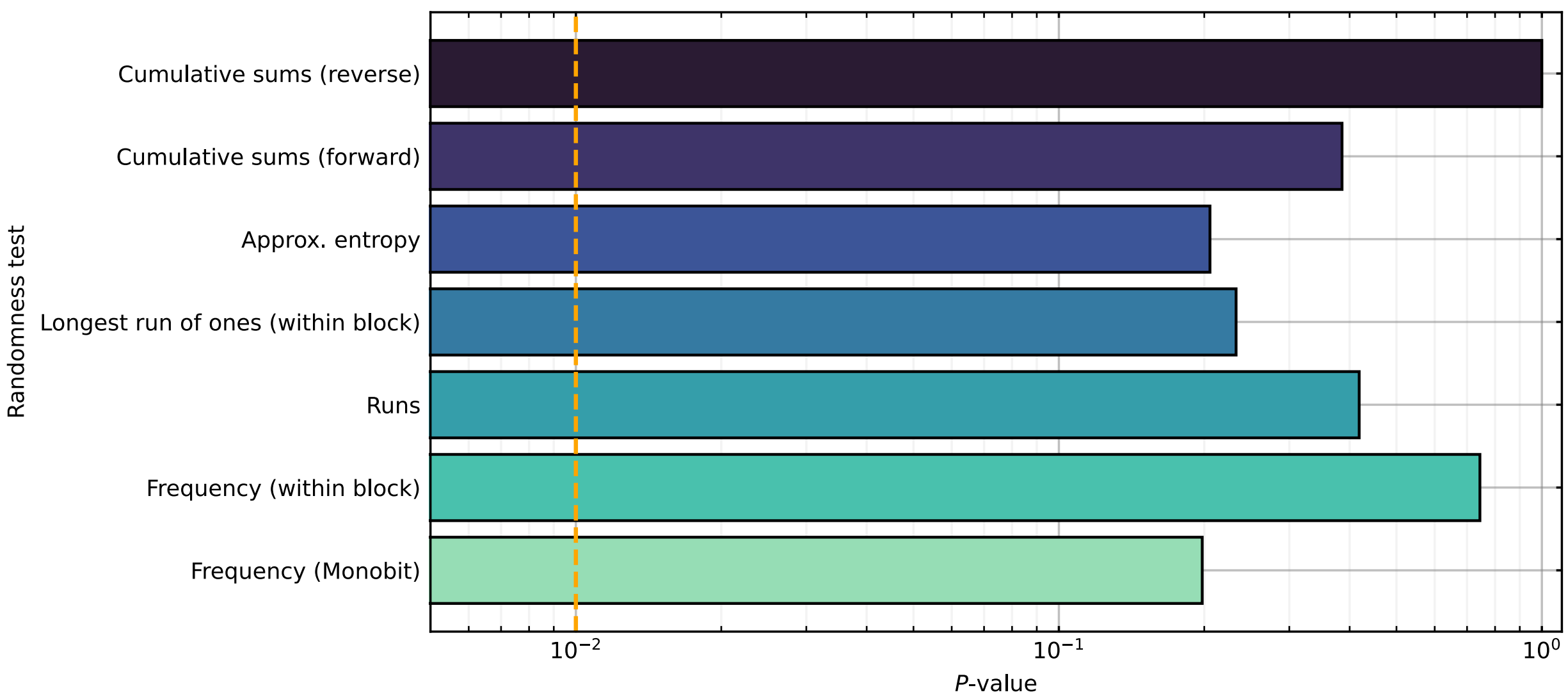

**Fig. S10: Statistical tests for randomness**

The bitstream generated from the phase-difference between bursts in multi-burst BSV was tested using the NIST Test Suite. The results of seven statistical tests are shown in the figure. The significance level used by the NIST Test Suite is $\alpha = 0.01$ which is denoted by the dashed orange line. For more information on the tests used, see the section on Towards PHz quantum random number generation in the Supplementary Discussion.